# Asymptotically-Optimal, Fast-Decodable, Full-Diversity STBCs

Lakshmi Prasad Natarajan and B. Sundar Rajan
Dept. of ECE, IISc, Bangalore 560012, India
Email: {nlp,bsrajan}@ece.iisc.ernet.in

*Abstract*—For a family/sequence of Space-Time Block Codes (STBCs) $\mathcal{C}_1, \mathcal{C}_2, \ldots$, with increasing number of transmit antennas $N_i$, with rates $R_i$ complex symbols per channel use (cspcu), $i = 1, 2, \ldots$, the *asymptotic normalized rate* is defined as $\lim_{i \to \infty} \frac{R_i}{N_i}$. A family of STBCs is said to be *asymptotically-good* if the asymptotic normalized rate is non-zero, i.e., when the rate scales as a non-zero fraction of the number of transmit antennas, and the family of STBCs is said to be *asymptotically-optimal* if the asymptotic normalized rate is 1, which is the maximum possible value. In this paper, we construct a new class of full-diversity STBCs that have the least maximum-likelihood (ML) decoding complexity among all known codes for any number of transmit antennas $N > 1$ and rates $R > 1$ cspcu. For a large set of $(R, N)$ pairs, the new codes have lower ML decoding complexity than the codes already available in the literature. Among the new codes, the class of full-rate codes ($R = N$) are asymptotically-optimal and fast-decodable, and for $N > 5$ have lower ML decoding complexity than all other families of asymptotically-optimal, fast-decodable, full-diversity STBCs available in the literature. The construction of the new STBCs is facilitated by the following further contributions of this paper: (i) For $g > 1$, we construct $g$-group ML-decodable codes with rates greater than one cspcu. These codes are asymptotically-good too. For $g > 2$, these are the first instances of $g$-group ML-decodable codes with rates greater than 1 cspcu presented in the literature. (ii) We construct a new class of fast-group-decodable codes (codes that combine the low ML decoding complexity properties of multigroup ML decodable codes and fast-decodable codes) for all even number of transmit antennas and rates $1 < R \leq 5/4$. (iii) Given a design with full-rank linear dispersion matrices, we show that a full-diversity STBC can be constructed from this design by encoding the real symbols independently using only regular PAM constellations[1].

## I. INTRODUCTION

Consider an $N$ transmit, $N_r$ receive antenna quasi-static, Rayleigh flat-faded MIMO channel $Y = XH + W$, where $X$ is the $T \times N$ transmitted codeword matrix, $H$ is the $N \times N_r$ channel matrix, $W$ is the $T \times N_r$ noise matrix and $Y$ is the $T \times N_r$ matrix of received signal. A *design* in $K$ real symbols $x_1, \ldots, x_K$ for $N$ transmit antennas is a matrix $\mathbf{X} = \sum_{i=1}^{K} x_i A_i$, where the *linear dispersion* or *weight matrices* $A_i \in \mathbb{C}^{T \times N}$, $i = 1, \ldots, K$, and the set $\{A_1, \ldots, A_K\}$ is linearly independent over $\mathbb{R}$. The rate of the design is $R = \frac{K}{2T}$ cspcu. We say that the design is of *high-rate* if $R > 1$ and *full-rate* if $R = N$. An STBC can be constructed from a design $\mathbf{X}$ by making the real symbols take values from a finite subset $\mathcal{A} \subset \mathbb{R}^K$. The finite set $\mathcal{A}$ is called the *signal set*. Denote the resulting STBC by $\mathcal{C}(\mathbf{X}, \mathcal{A})$, i.e., $\mathcal{C}(\mathbf{X}, \mathcal{A}) = \{\sum_{i=1}^{K} a_i A_i | [a_1, \ldots, a_K]^T \in \mathcal{A}\}$. For any non-empty subset $\Gamma \subset \{1, \ldots, K\}$, let $x(\Gamma) = \{x_i | i \in \Gamma\}$. An STBC $\mathcal{C}(\mathbf{X}, \mathcal{A})$ is said to be *$g$-group ML decodable*, $g > 1$, or *multigroup ML decodable* if there exists a partition $\Gamma_1, \ldots, \Gamma_g$ of the set $\{1, \ldots, K\}$ such that the symbol groups $x(\Gamma_1), \ldots, x(\Gamma_g)$ can be ML decoded independently of each other. If $|\Gamma_j| = K/g$ for all $j = 1, \ldots, g$, we say that the multigroup ML decodable code is *balanced*, else the code is said to be *unbalanced*. All the multigroup ML decodable codes considered in this paper are balanced. It is known [1], [2], [3], that if the weight matrices corresponding to symbols in different groups are Hurwitz-Radon orthogonal, i.e.,

$$A_p^H A_q + A_q^H A_p = \mathbf{0} \text{ for } p \in \Gamma_m, \ q \in \Gamma_n, \ 1 \leq m < n \leq g,$$

and if the symbol groups $x(\Gamma_1), \ldots, x(\Gamma_g)$ are encoded independently, then the STBC $\mathcal{C}(\mathbf{X}, \mathcal{A})$ is $g$-group ML decodable with respect to the partition $\Gamma_1, \ldots, \Gamma_g$. For such codes, if $M$ is the size of the underlying complex constellation, then the ML decoding complexity is only $\sum_{j=1}^{g} M^{\frac{|\Gamma_j|}{2}}$ instead of $M^{\sum_{j=1}^{g} \frac{|\Gamma_j|}{2}}$.

An STBC $\mathcal{C}(\mathbf{X}, \mathcal{A})$ is said to be *fast-decodable* [4] (FD) or *conditionally-$g$-group ML decodable*, $g > 1$, if there exists a non-empty subset $\Gamma \subsetneq \{1, \ldots, K\}$, such that for each of the values that $x(\Gamma^c)$ assumes, the remaining symbols $x(\Gamma)$ can be $g$-group ML decoded conditioned on the value of $x(\Gamma^c)$. Here, $\Gamma^c$ is the complement of $\Gamma$ in the universal set $\{1, \ldots, K\}$. If the conditional ML decoding groups in $\Gamma$ are balanced, then the ML decoding complexity is only $M^{\frac{|\Gamma^c|}{2}} \cdot gM^{\frac{|\Gamma|}{2g}}$ instead of $M^{\frac{|\Gamma|+|\Gamma^c|}{2}}$. An STBC $\mathcal{C}(\mathbf{X}, \mathcal{A})$ is said to be *fast-group-decodable* [5] (FGD) if the STBC is multigroup ML decodable with respect to a partition $\Gamma_1, \ldots, \Gamma_g$, $g > 1$ of $\{1, \ldots, K\}$ and atleast one group of symbols $x(\Gamma_j)$, $1 \leq j \leq g$, is fast-decodable. Such a code combines the low ML decoding complexity properties of FD and multigroup ML decodable codes.

Let $\{\mathcal{C}_1, \mathcal{C}_2, \ldots\}$ be an infinite family of STBCs for increasing number of transmit antennas $N_1, N_2, \ldots$ with rates $R_1, R_2, \ldots$ respectively.

*Definition 1:* The asymptotic rate of the family of codes $\{\mathcal{C}_1, \mathcal{C}_2, \ldots\}$ is defined to be the limit, $\lim_{i \to \infty} R_i$, if it exists.

---

[1]Part of the content of this manuscript has been accepted for presentation at IEEE GLOBECOM 2010, Miami, Florida. This work was supported partly by the DRDO-IISc program on Advanced Research in Mathematical Engineering through a research grant, and partly by the INAE Chair Professorship grant to B. S. Rajan

*Definition 2:* The asymptotic normalized rate of the family of codes $\{\mathcal{C}_1, \mathcal{C}_2, \ldots\}$ is defined to be the limit, $r = \lim_{i \to \infty} \frac{R_i}{N_i}$, if it exists. In other words, $r$ is the limit of the ratio of rate to full-rate of the given sequence of codes.

*Definition 3:* The family of codes $\{\mathcal{C}_1, \mathcal{C}_2, \ldots\}$ is said to be asymptotically-good if its asymptotic normalized rate, $r$ is non-zero, else the family of codes is said to be asymptotically-bad.

*Definition 4:* The family of codes $\{\mathcal{C}_1, \mathcal{C}_2, \ldots\}$ is said to be asymptotically-optimal if its asymptotic normalized rate, $r$ is equal to 1, the maximum possible value.

A family of codes with zero asymptotic rate is necessarily asymptotically-bad and a family with non-zero asymptotic rate may still be asymptotically-bad. In other words, an asymptotically-good family of codes has non-zero asymptotic rate, but the converse need not be true. The following examples as well as Table I illustrate this.

*Example 1: Families of multigroup ML decodable codes with zero asymptotic rate:* STBCs from orthogonal designs [6]–[8], satisfy $g = K$ and hence offer the least ML decoding complexity. In [8], maximal rate square complex orthogonal designs were given for number of antennas that are a power of 2. The rate for $2^m$ antennas is $\frac{m+1}{2^m}$. Single complex symbol ML-decodable codes, i.e., codes with $g = K/2$, were given in [2], [9], [10], [11]. For $2^m$ antennas, the rate of the single complex symbol decodable codes is $\frac{m}{2^{m-1}}$. Since the rates of orthogonal designs are low, quasi-orthogonal designs [12] were proposed with higher rate, although with higher ML decoding complexity. All these families of codes have asymptotic rate zero and hence are asymptotically-bad.

*Example 2: Families of multigroup ML decodable codes with non-zero asymptotic rate but are asymptotically-bad:* In [7], maximal rate orthogonal designs, not necessarily square, were constructed for arbitrary number of antennas. For number of antennas $N = 2m$ and $2m - 1$, $m > 1$, the codes constructed in [7] have a rate of $\frac{m+1}{2m}$. In [1], [2], the framework for multigroup ML-decodable codes was given. In [1], delay-optimal 4-group ML-decodable, rate-one codes were constructed for even number of antennas. In [3], codes for $2^m$ antennas with $g = 4$ and rate one were given based on Extended Clifford Algebras. Also in [3], Precoded Coordinate Interleaved Orthogonal Designs (PCIODs) having $g = 4$, $R = 1$ were given for even number of antennas in the setting of cooperative communications. Four-group ML-decodable, rate one codes for $2^m$ antennas were also given in [13]. Multigroup ML-decodable STBCs based on Clifford Unitary Weight Designs (CUWDs) were introduced in [14]. Low ML-decoding complexity STBCs that use Pauli matrices as the weight matrices were given in [15]. The $g$-group ML decodable codes from CUWDs [3] and from [15] can achieve a rate of $\frac{g}{2^{\lfloor \frac{g+1}{2} \rfloor}}$ cspcu, for arbitrary $g$ and number of transmit antennas $N = 2^m$, $m \geq \lceil \frac{g}{2} - 1 \rceil$.

*Example 3: All the known families of asymptotically-good multigroup ML decodable codes:* A 2-group ML-decodable, rate $5/4$ code for 4 transmit antennas was reported in [16]. In [17], 2-group ML-decodable, delay-optimal codes for $N = 2^m$ antennas, $m > 1$, were given based on Clifford Algebras. The code obtained in [17] for 4 antennas is equivalent to the code in [16]. The codes in [17] have a rate of $\frac{N}{4} + \frac{1}{N}$ cspcu. This class of codes is asymptotically-good, having $r = \frac{1}{4}$. Recently, unbalanced and balanced high-rate, 2-group ML-decodable codes were reported in [18] for any number of transmit antennas $N$. The class of balanced codes in [18] are not delay-optimal and they require $T \geq 2N$. These codes achieve a rate to full-rate ratio of $R/N = 1 - \frac{N}{T} + \frac{1}{NT}$. Thus, by increasing the delay $T$ appropriately with $N$, an asymptotic normalized rate of 1 can be achieved, and hence these codes are asymptotically-optimal. The family of codes in [17] and [18] are the only asymptotically-good multigroup ML-decodable codes reported in the literature so far.

*Example 4: All known families of asymptotically-optimal, full-diversity codes with low ML decoding complexity:* In [15], low ML decoding complexity, full-diversity STBCs with cubic shaping property [19] were constructed for $N = 2^m$ via codes over the finite field GF(4). The class of full-rate codes constructed in [15] for $2^m$, $m \geq 1$, antennas are conditionally 4-group ML decodable, information-lossless and constitute a family of asymptotically-optimal FD STBCs. Until this paper, the FD codes of [15] had the least known ML decoding complexity for $R > \frac{N}{4} + \frac{1}{N}$ and $N = 2^m$, $m \geq 1$. In [20], a family of full-diversity, full-rate STBCs called Threaded-Algebraic Space-Time (TAST) codes were constructed for all number of transmit antennas $N \geq 1$. These codes are conditionally 2-group ML decodable and hence constitute a family of asymptotically-optimal FD codes. The fast-decodability property of TAST codes, which we prove in Section V of this paper, was not reported in [20].

In [21], delay-optimal, full-rate and large coding gain codes were constructed from division algebras for arbitrary number of antennas. The asymptotic normalized rate of this family of codes is the maximum-possible value 1 and hence are asymptotically-optimal. However, these codes are 1-group ML-decodable, i.e., are not multigroup ML-decodable. The notion of fast-decodability was introduced in [4]. Rate 2 cspcu STBCs using designs with largest known coding gain were constructed for 2 and 4 antennas in [22]. These codes are fast-decodable as well. It was shown in [22] and [23] that the Golden Code [24] is fast-decodable. The Silver code, a rate-2 FD code for 2 transmit antennas, was found independently in [25] and [26]. In [27], FD codes, called Embedded Alamouti Space-Time codes, were constructed for $N = 2, 4, 6, 8$ and $R = 1, \ldots, N/2$ cspcu. A rate $3/2$ cspcu FD code for 4 antennas was constructed in [28], and in [29] a rate 2, 4 antenna FD code was constructed using Crossed Product Algebras. The 2 antenna code of [22] and the codes in [25]–[27], [29] have non-vanishing determinant property. FGD codes were introduced in [5], where a rate $17/8$ cspcu, 4 antenna FGD code was constructed. In [15], FGD codes were constructed for $2^m$ antennas and rates $1 < R \leq 5/4$.

The contributions and organization of this paper are as follows.

- We give a general procedure for constructing full-diversity, multigroup ML-decodable, delay-optimal codes. Using this procedure, we construct asymptotically-good, $g$-group ML decodable, delay-optimal, full-diversity STBCs for $g > 1$. For $g > 2$, the constructed class of STBCs contain the first instances of $g$-group ML decodable codes with rates greater than 1 cspcu and are the first instances of asymptotically-good $g$-group ML decodable codes reported in the literature. The class of 2-group ML decodable codes constructed in this paper match in terms of rate, the only other known class of asymptotically-good, delay-optimal 2-group ML-decodable codes, that were given in [17]. Moreover, the proposed class of codes is for all $N > 1$, whereas the codes in [17] are only for $N = 2^m$, $m > 1$. Specifically, we construct asymptotically-good, full-diversity, delay-optimal $g$-group ML decodable codes, $g > 1$, for $N = n2^{\lfloor \frac{g-1}{2} \rfloor}$ antennas, $n \geq g$, with rate $\frac{g}{N(1+\mathbf{1}\{g \text{ is even}\})} \left\lfloor \frac{N}{g2^{\lfloor \frac{g-1}{2} \rfloor}} \right\rfloor^2 + \frac{g^2-g}{2N}$ cspcu, where $\mathbf{1}\{\cdot\}$ is the indicator function (Section III).
- For $g \geq 3$, we propose $g$-group ML-decodable codes with the best-known asymptotic normalized rate. For $g = 2$, we construct codes that match the rates in [18] when the delays are same. Specifically, for $g > 1$, we propose non-delay-optimal codes with $N = n2^{\lfloor \frac{g-1}{2} \rfloor}$, $n \geq 1$, $T = gN$ and $R = \frac{N}{2^{g-1}} + \frac{g-1}{2N}$. The constructed codes have rate greater than one for $N \geq 2^{g-1}$ (Section III).
- We construct a new class of full-diversity, FGD codes for $N = 2m$, $m \geq 1$, antennas and rates $1 < R \leq 5/4$, which can be ML decoded with complexity $3M^{\frac{N}{4}(4R-3)-\frac{1}{2}}$. The new class of FGD codes match, in terms of ML decoding complexity, the class of FGD codes that were constructed in [15] for the same range of rates but for a smaller set of number of antennas $N = 2^m$, $m \geq 1$ (Section IV).
- Using the new asymptotically-good multigroup ML decodable codes and FGD codes, we construct a new class of delay-optimal, full-diversity STBCs that have the least ML decoding complexity among all known codes for all $N > 1$ and $R > 1$. For a large subset of $(R, N)$ pairs, the new codes have lower ML decoding complexity than the codes already available in the literature. Among the new codes, the class of full-rate codes are asymptotically-optimal and fast-decodable, and for $N > 5$, have lower ML decoding complexity than all other known families of asymptotically-optimal, full-diversity, FD codes (Table III in Section V gives a summary of comparison of ML decoding complexities of the new codes with other codes in the literature). In particular, whenever $N \geq 6$, $N$ not a power of 2 and $R > 1$, or $N \geq 6$, $N$ a power of 2 and $R > 3/2$, the new codes have lower ML decoding complexity than the codes available in the literature. For $N$ a power of 2 and $1 < R \leq 3/2$, the new codes have lower ML decoding complexity than already known codes when $N$ is sufficiently large. For all other $(R, N)$ pairs, the new codes match the existing codes in the literature in terms of ML decoding complexity (Section V).
- Given a delay-optimal design with full-rank weight matrices, we show that a full-diversity STBC can be constructed from this design by encoding the real symbols independently, using only regular PAM constellations (Section II).

Directions for future work are discussed in Section VI.

The new asymptotically-optimal FD codes of this paper for $N > 4$ antennas have lower ML decoding complexity than the asymptotically-optimal FD codes constructed in [15]. But, the codes in [15] possess additional desirable properties such as cubic shaping and information-losslessness which the new codes of this paper do not enjoy.

**Notation:** For a complex matrix $A$ the transpose, the conjugate and the conjugate-transpose are denoted by $A^T, \bar{A}$ and $A^H$ respectively. $A \otimes B$ is the Kronecker product of matrices $A$ and $B$. $I_n$ is the $n \times n$ identity matrix and $\mathbf{0}$ is the all zero matrix of appropriate dimension.. Cardinality of a set $\Gamma$ is denoted by $|\Gamma|$ and $i = \sqrt{-1}$. For a square matrix $A$, $det(A)$ is the determinant of $A$. For square matrices $A_1, \ldots, A_d$, $d \geq 1$, $diag(A_1, \ldots, A_d)$ denotes the square, block-diagonal matrix with $A_1, \ldots, A_d$ on the diagonal, in that order. $\mathbf{1}\{\cdot\}$ is the indicator function.

## II. FULL-DIVERSITY USING REGULAR PAM CONSTELLATIONS

In this section, we show that given a delay-optimal design $\mathbf{X}$ with full-rank weight matrices, we can create another design $\mathbf{X}'$, whose weight matrices are scaled versions of those of $\mathbf{X}$, such that the STBC obtained from $\mathbf{X}'$ by encoding the symbols with a regular PAM constellation is fully diverse. If the scaling coefficients belong to $\mathbb{R}$ this is equivalent to encoding each symbol of $\mathbf{X}$ with a regular PAM of possibly different minimum distance. In [15], it was shown that a full-diversity STBC can be obtained from a design with full-rank weight matrices by encoding the real symbols independently, but not necessarily using structured constellations, such as regular PAM. Further, the constellation used by each symbol in [15] is possibly different. In this case, in order to map the input information bits to symbols, a lookup table of length equal to the size of the constellation is required for each of the real symbols. However, each real symbol in the new design $\mathbf{X}'$ is independently encoded with the same regular PAM constellation, which can be effected by using a binary to decimal converter and thus leading to low encoding complexity.

Consider an $N \times N$ design $\mathbf{X} = \sum_{i=1}^{K} x_i A_i$, such that each $A_i$, $i = 1, \ldots, K$, is invertible. For any $\alpha = [\alpha_1, \alpha_2, \ldots, \alpha_K] \in \mathbb{C}^K$, let $\mathbf{X}'_\alpha = \sum_{i=1}^{K} x_i A'_i$, where $A'_i = \alpha_i A_i$, for $i = 1, \ldots, K$. For any subset $\mathcal{S} \subset \mathbb{C}$, let $\Delta \mathcal{S} = \{a - b | a, b \in \mathcal{S}\}$.

*Theorem 1:* Let $\mathfrak{B}$ be any infinite subset of $\mathbb{C}$ and $\mathcal{A}'$ be any finite subset of $\mathbb{R}$. Given an $N \times N$ design $\mathbf{X} = \sum_{i=1}^{K} x_i A_i$ with full-rank weight matrices $A_i$, there exists an $\alpha \in \mathfrak{B}^K$ such that the STBC $\mathcal{C}(\mathbf{X}'_\alpha, \mathcal{A}' \times \cdots \times \mathcal{A}')$ has full diversity.

*Proof:* Proof is via induction on $K$. We first prove the result for $K = 1$ and then prove the induction step.

When $K = 1$, we have $\mathbf{X} = x_1 A_1$. Choose $\alpha_1$ to be any non-zero element of $\mathfrak{B}$. Then $\mathbf{X}'_\alpha = x_1 \alpha_1 A_1$. Since $A_1$ is full-ranked, it is clear that the STBC $\mathcal{C}(\mathbf{X}'_\alpha, \mathcal{A}')$ is fully diverse.

We now prove the induction step. Let $K \geq 1$, $\tilde{\mathbf{X}} = \sum_{i=1}^{K} x_i A_i$ and $\tilde{\alpha} \in \mathfrak{B}^K$ be such that the STBC $\mathcal{C}_K = \mathcal{C}(\tilde{\mathbf{X}}'_{\tilde{\alpha}}, \mathcal{A}' \times \cdots \times \mathcal{A}')$ is fully diverse. Let $\mathbf{X} = \sum_{i=1}^{K+1} x_i A_i$, where $A_{K+1}$ is invertible. We now show that there exists an $\alpha_{K+1} \in \mathfrak{B}$ such that the STBC $\mathcal{C}_{K+1} = \mathcal{C}(\mathbf{X}'_\alpha, \mathcal{A}' \times \cdots \times \mathcal{A}')$ is fully diverse, where $\alpha = [\tilde{\alpha}^T \alpha_{K+1}]^T$. Consider the difference of any two codeword matrices in $\mathcal{C}_{K+1}$, say $\delta C$. Then, $\delta C = \delta C_a + \delta C_b$, where $\delta C_a$ is a difference of two codewords in $\mathcal{C}_K$, not necessarily different, and $\delta C_b \in \{\zeta \alpha_{K+1} A_{K+1} | \zeta \in \Delta \mathcal{A}'\}$. There are only a finite set of values that $\delta C_b$ can assume. The condition $\delta C \neq \mathbf{0}$ leads to the following cases.

1. $\delta C_a \neq \mathbf{0}$ and $\delta C_b = \mathbf{0}$: In this case, $\delta C = \delta C_a$, which is a codeword difference matrix of the fully diverse STBC $\mathcal{C}_K$. Hence $\delta C$ is invertible.

2. $\delta C_a \neq \mathbf{0}$ and $\delta C_b \neq \mathbf{0}$: In this case $\delta C = \delta C_a + \zeta \alpha_{K+1} A_{K+1}$. Let $f_{\delta C_a, \zeta}(z) = det(\delta C_a + \zeta z A_{K+1}) \in \mathbb{C}[z]$. $f_{\delta C_a, \zeta}(z)$ is not identically zero since $f_{\delta C_a, \zeta}(0) = det(\delta C_a) \neq 0$. For $\delta C$ to be full-rank, $\alpha_{K+1}$ should not be a zero of $f_{\delta C_a, \zeta}(z)$.

3. $\delta C_a = \mathbf{0}$ and $\delta C_b \neq \mathbf{0}$: In this case $\delta C = \zeta \alpha_{K+1} A_{K+1}$ where $\zeta \in \Delta \mathcal{A}' \setminus \{0\}$. $\delta C$ will be of full-rank as long as $\alpha_{K+1} \neq 0$.

From the above three cases it is clear that $\mathcal{C}_{K+1}$ will be fully-diverse as long as $\alpha_{K+1}$ is not a root of any of the polynomials $f_{\delta C_a, \zeta}(z)$ and $\alpha_{K+1} \neq 0$. Each of the polynomials is non-zero and hence has only a finite set of roots in $\mathbb{C}$. Further, there are only a finite number of polynomials $f_{\delta C_a, \zeta}(z)$. Hence, the set of values corresponding to $\alpha_{K+1}$ in $\mathbb{C}$ not leading to full diversity is finite. Thus, the set of values corresponding to $\alpha_{K+1}$ in $\mathfrak{B} \subset \mathbb{C}$ not leading to full diversity is also finite. Since $\mathfrak{B}$ is an infinite subset of $\mathbb{C}$, there exists an $\alpha_{K+1} \in \mathfrak{B}$, such that $\mathcal{C}_{K+1}$ is fully diverse. This proves the induction step. ∎

Let $\mathcal{A}_{Q-PAM} \subset \mathbb{R}$ denote the regular Q-ary PAM constellation with zero centroid and unit minimum Euclidean distance.

*Corollary 1:* Let $\mathfrak{B}$ be any infinite subset of $\mathbb{C}$ and $Q$ be any positive integer. Given an $N \times N$ design $\mathbf{X} = \sum_{i=1}^{K} x_i A_i$ with full-rank weight matrices $A_i$, there exists an $\alpha \in \mathfrak{B}^K$ such that the STBC $\mathcal{C}(\mathbf{X}'_\alpha, \mathcal{A}_{Q-PAM} \times \cdots \times \mathcal{A}_{Q-PAM})$ is fully diverse.

*Proof:* Follows from Theorem 1 on substituting $\mathcal{A}'$ with $\mathcal{A}_{Q-PAM}$. ∎

For $d > 0$, let $\mathcal{A}_{d,Q}$ denote the regular Q-ary PAM constellation with zero centroid and minimum Euclidean distance $d$.

*Corollary 2:* Given a positive integer $Q$ and an $N \times N$ design $\mathbf{X} = \sum_{i=1}^{K} x_i A_i$ with full-rank weight matrices $A_i$, there exist $d_i > 0$, $i = 1, \ldots, K$, such that the STBC $\mathcal{C}(\mathbf{X}, \mathcal{A}_{d_1,Q} \times \cdots \times \mathcal{A}_{d_K,Q})$ is fully diverse.

*Proof:* Let $\mathfrak{B} = \{\xi \in \mathbb{R} | \xi > 0\}$ and $\mathcal{A}' = \mathcal{A}_{Q-PAM}$. From Theorem 1, there exists $d = [d_1, \ldots, d_K]^T \in \mathfrak{B}^K$ such that the STBC $\mathcal{C}(\mathbf{X}'_\mathbf{d}, \mathcal{A}_{Q-PAM} \times \cdots \times \mathcal{A}_{Q-PAM})$ is fully diverse. But this STBC is same as $\mathcal{C}(\mathbf{X}, \mathcal{A}_{d_1,Q} \times \cdots \times \mathcal{A}_{d_K,Q})$. The desired result follows. ∎

We now give some examples with the help of STBCs already known in the literature.

*Example 5:* In [28], an FD STBC for $N = 4$ antennas and $R = \frac{3}{2}$ cspcu was constructed using full-rank weight matrices. Denote this design by $\mathbf{X} = \sum_{i=1}^{12} x_i A_i$. From Corollary 2, for a fixed $Q$, there exists a $[d_1, \ldots, d_{12}]^T \in \mathbb{R}^{12}$, such that $\mathcal{C}(\mathbf{X}, \mathcal{A}_{d_1,Q} \times \cdots \times \mathcal{A}_{d_{12},Q})$ is fully-diverse. In [28], it was shown that the choice $d_i = 1$, $i = 1, \ldots, 12$, provides full diversity for $Q = 2$ and $Q = 4$.

*Example 6:* The $2 \times 2$ Coordinate Interleaved Orthogonal Design (CIOD) [9] is constructed using the design $\begin{bmatrix} x_1 + ix_2 & 0 \\ 0 & x_3 + ix_4 \end{bmatrix}$. Here, $x_1 + ix_4$ and $x_3 + ix_2$ take values from square-QAM rotated by an angle $\varphi = \frac{1}{2} tan^{-1}(2) \ rad$. The STBC can be equivalently obtained from the design $\mathbf{X} = \sum_{i=1}^{4} x_i A_i$, where

$$A_1 = \begin{bmatrix} cos\varphi & 0 \\ 0 & isin\varphi \end{bmatrix}, \quad A_2 = \begin{bmatrix} -sin\varphi & 0 \\ 0 & icos\varphi \end{bmatrix},$$
$$A_3 = \begin{bmatrix} isin\varphi & 0 \\ 0 & cos\varphi \end{bmatrix}, \quad A_4 = \begin{bmatrix} icos\varphi & 0 \\ 0 & -sin\varphi \end{bmatrix}, \quad (1)$$

and by encoding each real symbol independently from $\mathcal{A}_{Q-PAM}$. Since each of the $A_i$, $i = 1, \ldots, 4$ is full rank, it follows from Corollary 2 that for a given value of $Q$, there exist $d_1, \ldots, d_4$ such that the signal set $\mathcal{A}_{d_1,Q} \times \cdots \times \mathcal{A}_{d_4,Q}$ leads to a full-diversity STBC.

*Proposition 1:* For the design (1), for any positive integer $Q$ and any $d = [d_1, d_2, d_3, d_4]^T \in \{\mathbb{Z}^4 | d_i > 0, i = 1, \ldots, 4\}$, the resulting STBC $\mathcal{C}(\mathbf{X}, \mathcal{A}_{d_1,Q} \times \cdots \times \mathcal{A}_{d_K,Q})$ has full diversity.

*Proof:* It is straightforward to show that the set of codeword difference matrices of the STBC $\mathcal{C}(\mathbf{X}, \mathcal{A}_{d_1,Q} \times \cdots \times \mathcal{A}_{d_K,Q})$ is a subset of the set of codeword difference matrices of the STBC $\mathcal{C}(\mathbf{X}, \mathcal{A}_{1,mQ} \times \cdots \times \mathcal{A}_{1,mQ})$, where $m = d_1 d_2 d_3 d_4$. It is sufficient to show that the STBC $\mathcal{C}(\mathbf{X}, \mathcal{A}_{1,mQ} \times \cdots \times \mathcal{A}_{1,mQ})$ has full diversity. Note that this STBC is same as the $2 \times 2$ CIOD whose complex symbols take values from rotated $(mQ)^2$-ary QAM. This code, however, is known to offer full diversity [9]. ∎

*Example 7:* In [22], a fast-decodable rate 2 code for 2 antennas was constructed using the design $\mathbf{X} = \sum_{i=1}^{8} x_i A_i$, where $A_1, \ldots, A_4$ are as given in (1) and

$$A_5 = \begin{bmatrix} 0 & isin\varphi \\ cos\varphi & 0 \end{bmatrix}, \quad A_6 = \begin{bmatrix} 0 & icos\varphi \\ -sin\varphi & 0 \end{bmatrix},$$
$$A_7 = \begin{bmatrix} 0 & cos\varphi \\ isin\varphi & 0 \end{bmatrix}, \quad A_8 = \begin{bmatrix} 0 & -sin\varphi \\ icos\varphi & 0 \end{bmatrix},$$

with $\varphi = \frac{1}{2} tan^{-1}(2) \ rad$. Note that all the weight matrices are of full rank. Let $\mathfrak{B} = \{e^{i\theta} | \theta \in [0, 2\pi)\}$ be the unit circle in the complex plane. From Corollary 1, there exist $\alpha_i$, $i = 1, \ldots, 8$ such that, $\alpha_i \in \mathfrak{B}$ and the STBC $\mathcal{C}(\mathbf{X}'_\alpha, \mathcal{A}_{Q-PAM} \times \cdots \times \mathcal{A}_{Q-PAM})$ has full diversity, for a given positive integer $Q$. The code obtained in [22] is the above STBC with $\alpha = [1, 1, 1, 1, e^{i\theta}, e^{i\theta}, e^{i\theta}, e^{i\theta}]^T$, where $\theta = \frac{\pi}{4}$. In [22], $\theta$ was optimized for diversity and coding gain. This code also has non-vanishing determinant property. This suggests that for a design with full-rank weight matrices, there may exist an $\alpha$ that leads to full diversity for any value of $Q$.

The ML decoding complexity of an STBC $\mathcal{C}(\mathbf{X}, \mathcal{A})$ is determined by both the design $\mathbf{X}$ and the signal set $\mathcal{A}$ [3]. For some $1 \le p < q \le K$, even if the weight matrices $A_p$ and $A_q$ are Hurwitz-Radon orthogonal, but if the choice of $\mathcal{A}$ is such that $x_p$ and $x_q$ have to be jointly encoded, then the two symbols need to be necessarily jointly ML decoded. The following definition can be used to denote the contribution of the choice of weight matrices towards ML decoding complexity of an STBC. Let, for any design $\mathbf{X} = \sum_{i=1}^{K} x_i A_i$ and non-empty subset $\Gamma \subset \{1, \ldots, K\}$, $\mathbf{X}_\Gamma$ denote the design $\sum_{i \in \Gamma} x_i A_i$.

*Definition 5 ( [15]):* Consider a design $\mathbf{X} = \sum_{i=1}^{K} x_i A_i$.

1) $\mathbf{X}$ is said to be $g$-group ML decodable if there exists a partition of $\{1, \ldots, K\}$ into $g$ non-empty subsets $\Gamma_1, \ldots, \Gamma_g$, such that

$$A_k^H A_l + A_l^H A_k = \mathbf{0} \text{ for } l \in \Gamma_i, k \in \Gamma_j, 1 \le i < j \le g.$$

2) $\mathbf{X}$ is said to be fast-decodable if there exists a non-empty subset $\Gamma \subsetneq \{1, \ldots, K\}$ such that the design $\mathbf{X}_\Gamma$ is $g$-group ML decodable for some $g > 1$.

3) $\mathbf{X}$ is said to be fast-group-decodable if $\mathbf{X}$ is $g$-group ML decodable with respect to the partition $\Gamma_1, \ldots, \Gamma_g$, and there exists an $i$, $1 \le i \le g$, such that $\mathbf{X}_{\Gamma_i}$ is fast-decodable.

If $\mathbf{X}$ is a multigroup ML decodable, FD or FGD design, then the STBC obtained from $\mathbf{X}$ by encoding the real symbols independently of each other is a multigroup ML decodable, FD or FGD STBC. All the new designs in this paper have full-rank weight matrices. From Corollary 2, full-diversity STBCs can be constructed from these designs by encoding the symbols independently with regular PAM constellations. In the remaining sections of the paper the focus is on constructing low ML decoding complexity designs and not on the construction of full-diversity achieving signal sets.

## III. NEW ASYMPTOTICALLY-GOOD, MULTIGROUP ML DECODABLE CODES

In this section, we first give a general procedure to construct multigroup ML decodable codes, and then using this procedure, we construct a new class of delay-optimal, full-diversity $g$-group ML decodable codes for $N = n2^{\lfloor \frac{g-1}{2} \rfloor}$ antennas with $g > 1$ and $n \ge g$, having asymptotic normalized rate $\frac{1}{g2^{g-1}}$. We also construct $g$-group ML decodable full-diversity STBCs for $N = n2^{\lfloor \frac{g-1}{2} \rfloor}$, $n \ge 1$, and delay $T = gN$ with asymptotic normalized rate $\frac{1}{2^{g-1}}$. For $g > 2$, these are the first instances of high-rate $g$-group ML decodable codes reported in the literature.

### A. General construction procedure

In this subsection, we give a general procedure to construct a full-diversity, delay-optimal $g$-group ML-decodable code, $g > 1$, starting from $g$ sets of matrices, $\mathcal{N}_l \subset \mathbb{C}^{N_l \times N_l}$, $l = 1, \ldots, g$, that satisfy the following two conditions for each $l = 1, \ldots, g$.

C1. The matrices in $\mathcal{N}_l$ are full-rank and linearly independent over $\mathbb{R}$,

C2. $\mathcal{N}_l$ can be partitioned as $\tilde{\mathcal{S}}_l \cup \mathcal{S}_l$, with $|\mathcal{S}_l| > 0$ and $|\tilde{\mathcal{S}}_l| = g - 1$. With $\tilde{\mathcal{S}}_l = \{A_{l,1}, \ldots, A_{l,l-1}, A_{l,l+1} \ldots, A_{l,g}\}$, the following hold

- $A_{l,j}^H B + B^H A_{l,j} = \mathbf{0}$ for all $A_{l,j} \in \tilde{\mathcal{S}}_l$, $B \in \mathcal{S}_l$ and
- $A_{l,j_1}^H A_{l,j_2} + A_{l,j_2}^H A_{l,j_1} = \mathbf{0}$ for $1 \le j_1 < j_2 \le g$ and $j_1, j_2 \ne l$.

The sets $\mathcal{N}_l$, $l = 1, \ldots, g$, need not be distinct. If the above conditions are satisfied by a set $\mathcal{N}_g$, then one can even use $\mathcal{N}_1 = \mathcal{N}_2 = \cdots = \mathcal{N}_g$.

We proceed by introducing functions $g_+$, $g_-$ and $f_l$ which are used to construct multigroup ML decodable codes starting from sets of matrices satisfying the conditions C1 and C2. Let $\mathcal{S}$ be a finite, non-empty subset of $\mathbb{C}^{N_1 \times N_1}$ consisting of full-rank matrices that are linearly independent over $\mathbb{R}$. Let $C \in \mathbb{C}^{N_2 \times N_2}$ be any full-rank matrix. Define

$$g_+(\mathcal{S}, C) = \{diag(A, C) | A \in \mathcal{S}\} \cup \{diag(A_0, -C)\} \text{ and}$$
$$g_-(\mathcal{S}, C) = \{diag(C, A) | A \in \mathcal{S}\} \cup \{diag(-C, A_0)\},$$

for some $A_0 \in \mathcal{S}$. Although $g_+(\mathcal{S}, C)$ and $g_-(\mathcal{S}, C)$ depend on the choice of $A_0$ in addition to $\mathcal{S}$ and $C$, the properties of $g_+(\mathcal{S}, C)$ and $g_-(\mathcal{S}, C)$ that this paper deals with, do not depend upon the particular choice of $A_0$. The results are valid for any choice of $A_0 \in \mathcal{S}$. Hence, for the sake of simplicity, we continue using this notation.

Both $g_+(\mathcal{S}, C)$ and $g_-(\mathcal{S}, C)$ are finite subsets of $\mathbb{C}^{(N_1+N_2) \times (N_1+N_2)}$ with cardinality $|\mathcal{S}|+1$ consisting of full-rank matrices. Further, if $C$ and all matrices in $\mathcal{S}$ are unitary, then all the matrices in $g_+(\mathcal{S}, C)$ and $g_-(\mathcal{S}, C)$ are unitary.

*Proposition 2:* The sets of matrices $g_+(\mathcal{S}, C)$ and $g_-(\mathcal{S}, C)$, are both linearly independent over $\mathbb{R}$.

*Proof:* Proof is given for $g_+(\mathcal{S}, C)$. Proof for $g_-(\mathcal{S}, C)$ is similar. Since the set $\mathcal{S}$ is linearly independent, it is clear that $\{diag(A, C) | A \in \mathcal{S}\}$ is linearly independent. It is enough to show that $diag(A_0, -C)$ does not belong to the subspace spanned by $\{diag(A, C) | A \in \mathcal{S}\}$ over $\mathbb{R}$. If $diag(A_0, -C)$ belongs to this subspace then, there exist a set of real numbers $\{\xi_A | A \in \mathcal{S}\}$ such that

$$diag(A_0, -C) = \sum_{A \in \mathcal{S}} \xi_A diag(A, C).$$

Since $\mathcal{S}$ is a linearly independent set, for all $A \ne A_0$, $\xi_A = 0$. Thus we have $diag(A_0, C) = \xi_{A_0} diag(A_0, -C)$. However, since both $A_0$ and $C$ are full-rank and hence non-zero, such a $\xi_{A_0}$ does not exist. Hence, $g_+(\mathcal{S}, C)$ is a linearly independent set. ∎

Let $d > 1$ be any integer. Let $N_1, \ldots, N_d$ be positive integers. Fix an $l \in \{1, \ldots, d\}$. For each $i \in \{1, \ldots, d\} \setminus \{l\}$, let $C_i \in \mathbb{C}^{N_i \times N_i}$ be full-rank. Let $\mathcal{S}$ be a finite, non-empty subset of $\mathbb{C}^{N_l \times N_l}$ consisting of full-rank matrices that are linearly independent over $\mathbb{R}$. Define $N = \sum_{i=1}^{d} N_i$. We construct a finite set, $f_l(\mathcal{S}, C_1, \ldots, C_{l-1}, C_{l+1}, \ldots, C_d) \subset \mathbb{C}^{N \times N}$, as follows. Let $\mathcal{S}_0 = \mathcal{S}$. For $i = 1, \ldots, l-1$, construct $\mathcal{S}_i$ recursively as: $\mathcal{S}_i = g_-(\mathcal{S}_{i-1}, C_i)$. For $i = l+1, \ldots, d$, construct $\mathcal{S}_{i-1}$ recursively as: $\mathcal{S}_{i-1} = g_+(\mathcal{S}_{i-2}, C_i)$. Finally, assign $f_l(\mathcal{S}, C_1, \ldots, C_{l-1}, C_{l+1}, \ldots, C_d) = \mathcal{S}_{d-1}$.

By the above construction and Proposition 2, we see that $f_l(\mathcal{S}, C_1, \ldots, C_{l-1}, C_{l+1}, \ldots, C_d)$ consists of full-rank, $\mathbb{R}$-linearly independent matrices. If the matrices $C_i$ and the matrices in $\mathcal{S}$ are unitary, the matrices in $f_l(\mathcal{S}, C_1, \ldots, C_{l-1}, C_{l+1}, \ldots, C_d)$ are also unitary. We note that $|f_l(\mathcal{S}, C_1, \ldots, C_{l-1}, C_{l+1}, \ldots, C_d)| = |\mathcal{S}| + d - 1$. It is clear that the matrices in $f_l(\mathcal{S}, C_1, \ldots, C_{l-1}, C_{l+1}, \ldots, C_d)$ have a block-diagonal structure, with $d$ blocks on the diagonal. For $i \in \{1, \ldots, d\} \setminus \{l\}$, the $i^{th}$ block is composed of $\pm C_i$. The $l^{th}$ block consists of matrices from $\mathcal{S} \cup -\mathcal{S}$.

For $l = 1, \ldots, g$, let $\mathcal{N}_l \subset \mathbb{C}^{N_l \times N_l}$ satisfy the conditions C1 and C2 given earlier in this subsection. For each $l = 1, \ldots, g$, define $\mathcal{M}_l = f_l(\mathcal{S}_l, A_{1,l}, \ldots, A_{l-1,l}, A_{l+1,l}, \ldots, A_{g,l})$. For each $l$, $\mathcal{M}_l \subset \mathbb{C}^{N \times N}$, where $N = \sum_{l=1}^{g} N_l$.

*Proposition 3:* $\mathcal{M} = \cup_{l=1}^{g} \mathcal{M}_l$ contains full-rank, $\mathbb{R}$-linearly independent matrices which satisfy

$$B_1^H B_2 + B_2^H B_1 = \mathbf{0} \text{ for } B_1 \in \mathcal{M}_{l_1}, B_2 \in \mathcal{M}_{l_2},$$

for $1 \leq l_1 < l_2 \leq g$.

*Proof:* The fact that the matrices are full-rank is clear. We now prove the Hurwitz-Radon orthogonality property.

Let $1 \leq l < k \leq g$ and $B_k \in \mathcal{M}_k$ and $B_l \in \mathcal{M}_l$. All the matrices in $\mathcal{M}$ have block diagonal structure, with the $j^{th}$ block as a matrix in $\mathbb{C}^{N_j \times N_j}$. Thus we have $B_l = diag(B_l[1], \ldots, B_l[g])$ and $B_k = diag(B_k[1], \ldots, B_k[g])$. Consider $j \notin \{l, k\}$. We have $B_l[j] = \pm A_{j,l}$ and $B_k[j] = \pm A_{j,k}$. Thus, $B_l[j]^H B_k[j] + B_k[j]^H B_l[j] = \mathbf{0}$. Now consider $j = l$. We have $B_l[j] \in \mathcal{S}_l \cup -\mathcal{S}_l$ and $B_k[j] = \pm A_{j,k}$. Thus, $B_l[j]^H B_k[j] + B_k[j]^H B_l[j] = \mathbf{0}$. Similar result holds for $j = k$ also. Thus for all $j = 1, \ldots, g$, we have $B_l[j]^H B_k[j] + B_k[j]^H B_l[j] = \mathbf{0}$. Hence $B_l^H B_k + B_k^H B_l = \mathbf{0}$.

We now show that the set $\mathcal{M}$ is linearly independent over $\mathbb{R}$. Let $span(\mathcal{M}_l)$ denote the subspace of $\mathbb{C}^{N \times N}$ spanned by the elements of $\mathcal{M}_l$ as a vector space over $\mathbb{R}$. From Proposition 2, it is clear that each $\mathcal{M}_l$ is a linearly independent set. It is therefore enough to show that $span(\mathcal{M}_l) \cap span(\mathcal{M}_k) = \{\mathbf{0}\}$ for $l \neq k$. In order to prove this, let $B \in span(\mathcal{M}_l) \cap span(\mathcal{M}_k)$. Therefore for some set of real numbers $\{\xi_A | A \in \mathcal{M}_k\}$, $B = \sum_{A \in \mathcal{M}_k} \xi_A A$. Since $B \in span(\mathcal{M}_l)$ we have, for all $A \in \mathcal{M}_k$, $B^H A + A^H B = \mathbf{0}$. Hence, $B^H B + B^H B = \mathbf{0}$. Thus, $B = \mathbf{0}$. ∎

We construct an STBC $\mathcal{C} \subset \mathbb{C}^{N \times N}$, by using the matrices in $\mathcal{M}$ as the linear dispersion matrices of a design $\mathbf{X}$. Let $\mathbf{X} = \sum_{B \in \mathcal{M}} x_B B$, where the real symbols $x_B$ are indexed by the elements of $\mathcal{M}$. From Corollary 2, the design $\mathbf{X}$, obtained via the proposed construction method, can be combined with appropriate regular PAM constellations to obtain a full-diversity, $g$-group ML-decodable STBC $\mathcal{C}$. For $l = 1, \ldots, g$, the $l^{th}$ decoding group is $\{x_A | A \in \mathcal{M}_l\}$. The rate of the code in cspcu is $R = \frac{|\mathcal{M}|}{2N}$. Since $|\mathcal{M}_l| = |\mathcal{S}_l| + g - 1$, we have

$$R = \frac{\sum_{l=1}^{g} (|\mathcal{S}_l| + g - 1)}{2 \sum_{l=1}^{g} N_l}. \tag{2}$$

The following proposition will be useful when we are constructing high-rate multigroup ML-decodable codes in Section III-B.

*Proposition 4:* Let $k \geq 1$. There exists an explicitly constructable set $\mathcal{H}_k \subset \mathbb{C}^{k \times k}$ consisting of $k^2$ matrices that are unitary, Hermitian and linearly independent over $\mathbb{R}$.

*Proof:* For $k = 1$, $\mathcal{H}_1 = \{[1]\}$ satisfies all the properties in the hypothesis. Let us assume that $k \geq 2$.

For a matrix $A$, let $A(n, m)$ denote the entry in the $n^{th}$ row and $m^{th}$ column of $A$. For $n = 1, \ldots, k$, define $Z_n \in \mathbb{C}^{k \times k}$ as follows. $Z_1 = diag(1, 1, 1, \ldots, 1)$, $Z_2 = diag(1, -1, 1, \ldots, 1)$, $Z_3 = diag(1, 1, -1, \ldots, 1)$, ..., $Z_k = diag(1, 1, 1, \ldots, -1)$. For $n = 1, \ldots, k-1$ and $m > n$ define $V_{n,m} \in \mathbb{C}^{k \times k}$ as follows.

$$V_{n,m}(n, m) = V_{n,m}(m, n) = 1, \; V_{n,m}(l, l) = 1 \text{ for } l \neq n, m.$$

All other entries of $V_{n,m}$ are set to zero. For $n = 1, \ldots, k-1$ and $m > n$ define $W_{n,m} \in \mathbb{C}^{k \times k}$ as follows.

$$W_{n,m}(n, m) = -W_{n,m}(m, n) = i, W_{n,m}(l, l) = 1, l \neq n, m.$$

All other entries of $W_{n,m}$ are set to zero. We now prove that $\mathcal{H}_k = \{Z_n\} \cup \{V_{n,m}\} \cup \{W_{n,m}\}$ satisfies all the properties in the hypothesis. It is straightforward to see that every matrix in $\mathcal{H}_k$ is unitary and Hermitian. Further, $|\{Z_n\}| = k$, $|\{V_{n,m}\}| = \frac{k(k-1)}{2}$ and $|\{W_{n,m}\}| = \frac{k(k-1)}{2}$. Thus, $|\mathcal{H}_k| = k^2$.

It only remains to show that the set $\mathcal{H}_k$ is linearly independent over $\mathbb{R}$. For $n = 1, \ldots, k$ and $m > n$, let $\zeta_n, \alpha_{n,m}, \beta_{n,m}$ be real numbers such that

$$\sum_{n=1}^{k} \zeta_n Z_n + \sum_{n=1}^{k-1} \sum_{m > n} (\alpha_{n,m} V_{n,m} + \beta_{n,m} W_{n,m}) = \mathbf{0}.$$

Then, for each $1 \leq p, q \leq n$ we have

$$\sum_{n=1}^{k} \sum_{m > n} (\alpha_{n,m} V_{n,m}(p, q) + \beta_{n,m} W_{n,m}(p, q))$$

$$+ \sum_{n=1}^{k} \zeta_n Z_n(p, q) = 0.$$

For $p = 1, \ldots, k-1$ and $q > p$, we have $Z_n(p, q) = 0$ for all $n$, $V_{n,m}(p, q) = \delta_{n,p} \delta_{m,q}$ and $W_{n,m}(p, q) = i\delta_{n,p}\delta_{m,q}$. Thus, $\alpha_{p,q} + i\beta_{p,q} = 0$. Hence, we have $\alpha_{p,q} = \beta_{p,q} = 0$. Since the above sum reduces to $\sum_{n=1}^{k} \zeta_n Z_n$, it only remains to show that $\zeta_n = 0$ for all $n$. $\sum_{n=1}^{k} \zeta_n Z_n = \mathbf{0}$ is equivalent to the

proposition that $(\zeta_1, \ldots, \zeta_k)^T$ is in the null-space of the matrix

$$\begin{bmatrix} 1 & 1 & 1 & \cdots & 1 \\ 1 & -1 & 1 & \cdots & 1 \\ 1 & 1 & -1 & \cdots & 1 \\ \vdots & \vdots & \vdots & & \vdots \\ 1 & 1 & 1 & \cdots & -1 \end{bmatrix} = \mathbf{11}^T - diag(0, 2, \ldots, 2),$$

where $\mathbf{1}$ is the all ones vector. Thus, we have

$$\mathbf{1} \left( \sum_{n=1}^{k} \zeta_n \right) = diag(0, 2\zeta_2, \ldots, 2\zeta_k).$$

Equating the components of the vectors on either side of the equation, we have $0 = \sum_{n=1}^{k} \zeta_n = 2\zeta_m$ for $m > 1$. Thus, $0 = \sum_{n=1}^{k} \zeta_n = \zeta_1$. This completes the proof. ∎

*Example 8:* The set $\mathcal{H}_2 \subset \mathbb{C}^{2 \times 2}$ consists of the following 4 matrices:

$$\begin{bmatrix} 1 & 0 \\ 0 & 1 \end{bmatrix}, \begin{bmatrix} 1 & 0 \\ 0 & -1 \end{bmatrix}, \begin{bmatrix} 0 & 1 \\ 1 & 0 \end{bmatrix} \text{ and } \begin{bmatrix} 0 & i \\ -i & 0 \end{bmatrix}.$$

*Example 9:* The set $\mathcal{H}_3 \subset \mathbb{C}^{3 \times 3}$ consists of the following 9 matrices:

$$\begin{bmatrix} 1 & 0 & 0 \\ 0 & 1 & 0 \\ 0 & 0 & 1 \end{bmatrix}, \begin{bmatrix} 1 & 0 & 0 \\ 0 & -1 & 0 \\ 0 & 0 & 1 \end{bmatrix}, \begin{bmatrix} 1 & 0 & 0 \\ 0 & 1 & 0 \\ 0 & 0 & -1 \end{bmatrix}, \begin{bmatrix} 0 & 1 & 0 \\ 1 & 0 & 0 \\ 0 & 0 & 1 \end{bmatrix},$$

$$\begin{bmatrix} 0 & i & 0 \\ -i & 0 & 0 \\ 0 & 0 & 1 \end{bmatrix}, \begin{bmatrix} 0 & 0 & 1 \\ 0 & 1 & 0 \\ 1 & 0 & 0 \end{bmatrix}, \begin{bmatrix} 0 & 0 & i \\ 0 & 1 & 0 \\ -i & 0 & 0 \end{bmatrix},$$

$$\begin{bmatrix} 1 & 0 & 0 \\ 0 & 0 & 1 \\ 0 & 1 & 0 \end{bmatrix} \text{ and } \begin{bmatrix} 1 & 0 & 0 \\ 0 & 0 & i \\ 0 & -i & 0 \end{bmatrix}.$$

*B. Asymptotically-good, delay-optimal multigroup ML decodable codes*

In this subsection we construct $\{\mathcal{N}_l | l = 1, \ldots, g\}$ leading to a family of delay-optimal, asymptotically-good, $g$-group ML-decodable codes, for $g > 1$.

*Theorem 2 ( [8]):* The maximum rate of a square orthogonal design in $K$ complex symbols is $\frac{K}{2^{K-1}}$ cspcu.

From Theorem 2, we see that, there exists a square orthogonal design of dimension $2^{K-1} \times 2^{K-1}$ in $K$ complex symbols, i.e., in $2K$ real symbols. The linear dispersion matrices of a square orthogonal design are all unitary and are pairwise Hurwitz-Radon orthogonal. Thus, for $m \geq 0$, there exist $2(m+1)$ unitary, $\mathbb{R}$-linearly independent, pairwise Hurwitz-Radon orthogonal $2^m \times 2^m$ complex matrices. Let us denote this set of matrices by $\mathcal{O}_m = \{U_1, \ldots, U_{2m+2}\}$.

Let $g > 1$ be any integer. Let $N = n2^{\lfloor \frac{g-1}{2} \rfloor}$ and $n \geq g$. Then, $\frac{n}{g} = p + \epsilon$, where $p \geq 1$ is an integer and $\epsilon \in [0, 1)$. Then, $n = gp + g\epsilon$, where $gp$ is a positive integer and $t = g\epsilon \in \{0, 1, \ldots, g-1\}$. Assign $N_1 = N_2 = \cdots = N_t = (p+1)2^{\lfloor \frac{g-1}{2} \rfloor}$ and $N_{t+1} = \cdots = N_g = p2^{\lfloor \frac{g-1}{2} \rfloor}$. Clearly, $\sum_{l=1}^{g} N_l = N$. Let $\mathcal{H}_p \subset \mathbb{C}^{p \times p}$ and $\mathcal{H}_{p+1} \subset \mathbb{C}^{p+1 \times p+1}$ be the sets of matrices promised by Proposition 4. Let $\bar{\mathcal{H}}_{p+1}$ be any subset of $\mathcal{H}_{p+1}$ of cardinality $p^2$. The construction procedure is given in two cases below.

*Case 1, g is even:* Here, $g = 2(m+1)$ for some $m \geq 0$ and $m = \lfloor \frac{g-1}{2} \rfloor$. With $U_i$ as the weight matrices of the $2^m \times 2^m$ maximal rate square complex orthogonal design, for $l = 1, \ldots, t$, define:

$$\tilde{\mathcal{S}}_l = \{I_{p+1} \otimes U_i | i = 1, \ldots, 2m+1\} \text{ and}$$
$$\mathcal{S}_l = \{A \otimes U_{2m+2} | A \in \bar{\mathcal{H}}_{p+1}\}.$$

For $l = t+1, \ldots, g$, define:

$$\tilde{\mathcal{S}}_l = \{I_p \otimes U_i | i = 1, \ldots, 2m+1\} \text{ and}$$
$$\mathcal{S}_l = \{A \otimes U_{2m+2} | A \in \mathcal{H}_p\}.$$

Assign $\mathcal{N}_l = \mathcal{S}_l \cup \tilde{\mathcal{S}}_l$, $l = 1, \ldots, g$. It is straightforward to show that the sets of matrices $\{\mathcal{N}_l | l = 1, \ldots, g\}$ satisfy the conditions C1 and C2 of the construction procedure in Section III-A. Number of real symbols in each group of the design resulting from the construction procedure in Section III-A applied to the sets of matrices $\{\mathcal{N}_l\}$ is $p^2 + g - 1$. The rate of the code is $\frac{g(p^2+g-1)}{2N}$ which is the same as

$$R = \frac{g}{2N} \left\lfloor \frac{N}{g2^{\lfloor \frac{g-1}{2} \rfloor}} \right\rfloor^2 + \frac{g^2 - g}{2N} \; cspcu. \quad (3)$$

*Case 2, g is odd:* In this case, $g = 2m+1$, for some $m \geq 1$. With $U_i$ as the weight matrices of the $2^m \times 2^m$ maximal rate square complex orthogonal design, for $l = 1, \ldots, t$, define:

$$\tilde{\mathcal{S}}_l = \{I_{p+1} \otimes U_i | i = 1, \ldots, 2m\} \text{ and}$$
$$\mathcal{S}_l = \{A \otimes B | A \in \bar{\mathcal{H}}_{p+1}, B \in \{U_{2m+1}, U_{2m+2}\}\}.$$

For $l = t+1, \ldots, g$, define:

$$\tilde{\mathcal{S}}_l = \{I_p \otimes U_i | i = 1, \ldots, 2m\} \text{ and}$$
$$\mathcal{S}_l = \{A \otimes B | A \in \mathcal{H}_p, B \in \{U_{2m+1}, U_{2m+2}\}\}.$$

Assign $\mathcal{N}_l = \mathcal{S}_l \cup \tilde{\mathcal{S}}_l$, $l = 1, \ldots, g$. Again, it is straightforward to show that the sets of matrices $\{\mathcal{N}_l | l = 1, \ldots, g\}$ satisfy the conditions C1 and C2 of the construction procedure in Section III-A. Number of real symbols in each group of the design resulting from the construction procedure in Section III-A applied to the sets of matrices $\{\mathcal{N}_l\}$ is $2p^2 + g - 1$. The rate of the code is $\frac{g(2p^2+g-1)}{2N}$ which is the same as

$$R = \frac{g}{N} \left\lfloor \frac{N}{g2^{\lfloor \frac{g-1}{2} \rfloor}} \right\rfloor^2 + \frac{g^2 - g}{2N} \; cspcu. \quad (4)$$

From (3), (4) and Corollary 2, for any $n \geq g$, we have a $g$-group ML-decodable, delay-optimal, full-diversity STBC $\mathcal{C}$ for number of transmit antennas $N = n2^{\lfloor \frac{g-1}{2} \rfloor}$, with rate

$$R(g, N) = \frac{g}{N(1 + \mathbf{1}\{g \text{ is even}\})} \left\lfloor \frac{N}{g2^{\lfloor \frac{g-1}{2} \rfloor}} \right\rfloor^2 + \frac{g^2 - g}{2N}. \quad (5)$$

Every codeword matrix of $\mathcal{C}$ is block-diagonal, with $g$ square complex matrices on the diagonal. The first $t$ matrices are of dimension $(p+1)2^{\lfloor \frac{g-1}{2} \rfloor}$ and the remaining $g - t$ matrices are of dimension $p2^{\lfloor \frac{g-1}{2} \rfloor}$. When $N = pg2^{\lfloor \frac{g-1}{2} \rfloor}$, $p \geq 1$, i.e., when $n = pg$, we have $\epsilon = t = 0$. Thus, every codeword in $\mathcal{C}$ is block-diagonal with $g$ complex matrices on the diagonal, each of dimension $p2^{\lfloor \frac{g-1}{2} \rfloor} \times p2^{\lfloor \frac{g-1}{2} \rfloor}$.

*Example 10:* Consider $N = 6$ and $g = 2$. In this case, $n = 6$ and is a multiple of $g$. Hence, $p = 3$ and $t = 0$. Further, $m = \lfloor \frac{g-1}{2} \rfloor = 0$ and the maximal rate square orthogonal design of size $2^m = 1$ has weight matrices $\{[1], [i]\}$. Thus, $\mathcal{S}_1 = \mathcal{S}_2 = i\mathcal{H}_3$ and $\tilde{\mathcal{S}}_1 = \tilde{\mathcal{S}}_2 = \{I_3\}$ where $\mathcal{H}_3$ is given in Example 9. The two groups of weight matrices are $\mathcal{M}_1 = f_1(i\mathcal{H}_3, I_3)$ and $\mathcal{M}_2 = f_2(i\mathcal{H}_3, I_3)$. Since $|\mathcal{H}_3| = 9$, the number of real symbols in each of the two groups is 10. The rate of the resulting design is $5/3$ cspcu. Let $\{x_1, \ldots, x_{10}\}$ and $\{x_{11}, \ldots, x_{20}\}$ be the symbols associated with the weight matrices in $\mathcal{M}_1$ and $\mathcal{M}_2$ respectively. The resulting 2-group ML-decodable code is given in (6) at the top of the next page, where the auxiliary variables in the design, $z_j, j = 1, \ldots, 8$, are

$$z_1 = x_1 + x_2 + x_3 + x_8 + x_9 + x_{10},$$
$$z_2 = x_1 - x_2 + x_3 + x_6 + x_7 + x_{10},$$
$$z_3 = x_1 + x_2 - x_3 + x_4 + x_5 + x_{10},$$
$$z_4 = x_{11} + x_{12} + x_{13} + x_{18} + x_{19} + x_{20},$$
$$z_5 = x_{11} - x_{12} + x_{13} + x_{16} + x_{17} + x_{20},$$
$$z_6 = x_{11} + x_{12} - x_{13} + x_{14} + x_{15} + x_{20},$$
$$z_7 = \sum_{j=1}^{10} x_j, \text{ and } z_8 = \sum_{j=11}^{20} x_j.$$

*Example 11:* Let $N = 12$ and $g = 3$. We see that $n = 6$ is a multiple of $g$. Thus, $\epsilon = t = 0$, $p = \frac{n}{g} = 2$ and $m = \lfloor \frac{g-1}{2} \rfloor = 1$. Thus, we are concerned with square orthogonal design of size $2^m = 2$. This is exactly the Alamouti code [30], whose weight matrices are $U_1 = \begin{bmatrix} 1 & 0 \\ 0 & 1 \end{bmatrix}$, $U_2 = \begin{bmatrix} i & 0 \\ 0 & -i \end{bmatrix}$, $U_3 = \begin{bmatrix} 0 & i \\ i & 0 \end{bmatrix}$ and $U_4 = \begin{bmatrix} 0 & -1 \\ 1 & 0 \end{bmatrix}$. Let $A_1 = I_2 \otimes U_1$ and $A_2 = I_2 \otimes U_2$. With $\mathcal{H}_2$ as given in Example 8, let $\mathcal{S} = \{A \otimes B | A \in \mathcal{H}_2, B \in \{U_3, U_4\}\}$. Let $\mathcal{N}_1 = \mathcal{N}_2 = \mathcal{N}_3 = \mathcal{S} \cup \{A_1, A_2\}$. The three groups of weight matrices are then given by $\mathcal{M}_1 = f_1(\mathcal{S}, A_1, A_2)$, $\mathcal{M}_2 = f_2(\mathcal{S}, A_2, A_1)$ and $\mathcal{M}_3 = f_3(\mathcal{S}, A_1, A_2)$. Each group is composed of $2p^2 + g - 1 = 10$ real symbols. The rate of this design is $5/4$ cspcu. Associate the real symbols $\{x_1, \ldots, x_{10}\}$, $\{x_{11}, \ldots, x_{20}\}$ and $\{x_{21}, \ldots, x_{30}\}$ with the weight matrices in $\mathcal{M}_1$, $\mathcal{M}_2$ and $\mathcal{M}_3$ respectively. For $k = 1, 2, 3$, define the following variables.

$$v_k = x_{4+10(k-1)} + x_{7+10(k-1)} + i(x_{3+10(k-1)} - x_{8+10(k-1)}),$$
$$y_k = x_{4+10(k-1)} - x_{7+10(k-1)} + i(x_{3+10(k-1)} + x_{8+10(k-1)}),$$

$$u_k = x_{5+10(k-1)} + x_{6+10(k-1)} + $$
$$i(x_{1+10(k-1)} + x_{2+10(k-1)} + x_{9+10(k-1)} + x_{10k}), \text{ and}$$

$$z_k = x_{5+10(k-1)} - x_{6+10(k-1)} + $$
$$i(x_{1+10(k-1)} - x_{2+10(k-1)} + x_{9+10(k-1)} + x_{10k}).$$

Also define

$$w_1 = \sum_{j=21}^{28} x_j - x_{29} + x_{30} + i \left( \sum_{j=11}^{18} x_j + x_{19} - x_{20} \right),$$
$$w_2 = \sum_{j=1}^{8} x_j - x_9 + x_{10} + i \left( \sum_{j=21}^{28} x_j + x_{29} - x_{30} \right) \&$$
$$w_3 = \sum_{j=11}^{18} x_j - x_{19} + x_{20} + i \left( \sum_{j=1}^{8} x_j + x_9 - x_{10} \right).$$

The resulting design is given in (7) at the top of the next page.

*Example 12:* Consider $N = 20$ and $g = 3$. $N$ is a multiple of $2^{\lfloor \frac{g-1}{2} \rfloor} = 2$. A full-diversity, 3-group ML decodable code can be constructed for $N = 20$ using the procedure given in this subsection. From (5), this new code has rate $R = 3/2$ cspcu.

We now show that the new class of $g$-group ML decodable codes, $g > 1$, constructed in this subsection has asymptotic-normalized-rate $\frac{1}{g2^{g-1}}$ and hence is asymptotically-good. Thus, by increasing $N$, we get $g$-group ML decodable codes with rates greater than 1 cspcu for any $g > 1$.

*Proposition 5:* Let $g > 1$. The asymptotic normalized rate of the family of $g$-group ML decodable STBCs constructed in this subsection for number of antennas $N \in \{n2^{\lfloor \frac{g-1}{2} \rfloor} | n \in \mathbb{Z}, n \geq g\}$ is $\frac{1}{g2^{g-1}}$.

*Proof:* We prove the result for the case when $g$ is odd. The proof for even values of $g$ is similar. Consider the ratio of rate (4) to number of antennas $N$,

$$\frac{R}{N} = \frac{g^2 - g}{2N^2} + \frac{g}{N^2} \left\lfloor \frac{N}{g2^{\frac{g-1}{2}}} \right\rfloor^2.$$

Since $N = n2^{\frac{g-1}{2}}$, we have $\left\lfloor \frac{N}{g2^{\frac{g-1}{2}}} \right\rfloor = \left\lfloor \frac{n}{g} \right\rfloor = \frac{n}{g} - \epsilon_N$, $\epsilon_N \in [0, 1)$. Therefore,

$$\frac{R}{N} = \frac{g^2 - g}{2n^2 2^{g-1}} + \frac{g}{n^2 2^{g-1}} \left( \frac{n}{g} - \epsilon_N \right)^2$$
$$= \frac{g^2 - g}{n^2 2^g} + \frac{g}{n^2 2^{g-1}} \left( \frac{n^2}{g^2} - \frac{2n\epsilon_N}{g} + \epsilon_N^2 \right)$$
$$= \frac{g^2 - g}{n^2 2^g} + \frac{1}{g2^{g-1}} - \frac{2\epsilon_N}{ng2^{g-1}} + \frac{g\epsilon_N^2}{n^2 2^{g-1}}. \quad (8)$$

The limit $N \to \infty$ corresponds to the limit $n \to \infty$. When $n \to \infty$, in the summation (8), only the second term does not vanish. This completes the proof. ∎

*C. Asymptotically-good, non-delay-optimal, multigroup ML-decodable codes*

In this subsection, we construct non-delay-optimal, full-diversity, $g$-group ML-decodable codes for $g > 1$, using the codes in Section IV, leading to higher asymptotic normalized rates.

Let $A$ be a square, block-diagonal, complex matrix with $g$ square matrices $\{C_1, \ldots, C_g\}$, each of dimension $k \times k$ on its diagonal. Define $\phi(A, g) = [C_1^T, \ldots, C_g^T]^T$. Note that $\phi(A, g)$ is of dimension $gk \times k$. If $A_1, \ldots, A_K$ are $K$ such block-diagonal matrices that are linearly independent over $\mathbb{R}$ then, it is clear that, $\phi(A_1, g), \ldots, \phi(A_K, g)$ are also linearly independent over $\mathbb{R}$.

Let $p \geq 1$ and $N = pg2^{\lfloor \frac{g-1}{2} \rfloor}$. Let $\mathcal{C}$ be a $g$-group ML-decodable, delay optimal STBC obtained from the construction in Section III-B, using a design $\mathbf{X} = \sum_{i=1}^{K} x_i A_i$ and a full-diversity achieving signal set $\mathcal{A} \subset \mathbb{R}^K$. That is, $\mathcal{C} = \{\sum_{i=1}^{K} \xi_i A_i | (\xi_1, \ldots, \xi_K) \in \mathcal{A}\}$. Every codeword in $\mathcal{C}$ is block-diagonal with $g$ complex matrices on the diagonal, each of dimension $p2^{\lfloor \frac{g-1}{2} \rfloor} \times p2^{\lfloor \frac{g-1}{2} \rfloor}$. From (5),

$$\mathbf{X_{6\times 6}} = \begin{bmatrix} z_8+iz_1 & -x_5+ix_4 & -x_7+ix_6 & 0 & 0 & 0 \\ x_5+ix_4 & z_8+iz_2 & -x_9+ix_8 & 0 & 0 & 0 \\ x_7+ix_6 & x_9+ix_8 & z_8+iz_3 & 0 & 0 & 0 \\ 0 & 0 & 0 & z_7+iz_4 & -x_{15}+ix_{14} & -x_{17}+ix_{16} \\ 0 & 0 & 0 & x_{15}+ix_{14} & z_7+iz_5 & -x_{19}+ix_{18} \\ 0 & 0 & 0 & x_{17}+ix_{16} & x_{19}+ix_{18} & z_7+iz_6 \end{bmatrix} \quad (6)$$

$$\mathbf{X_{12\times 12}} = diag\left(\begin{bmatrix} w_1 & -u_1^* & 0 & -v_1^* \\ u_1 & w_1^* & -y_1^* & 0 \\ 0 & y_1 & w_1 & -z_1^* \\ v_1 & 0 & z_1 & w_1^* \end{bmatrix}, \begin{bmatrix} w_2 & -u_2^* & 0 & -v_2^* \\ u_2 & w_2^* & -y_2^* & 0 \\ 0 & y_2 & w_2 & -z_2^* \\ v_2 & 0 & z_2 & w_2^* \end{bmatrix}, \begin{bmatrix} w_3 & -u_3^* & 0 & -v_3^* \\ u_3 & w_3^* & -y_3^* & 0 \\ 0 & y_3 & w_3 & -z_3^* \\ v_3 & 0 & z_3 & w_3^* \end{bmatrix}\right) \quad (7)$$

we have $K = \frac{2N^2}{g2^{g-1}} + g^2 - g$. Let $\Gamma_1, \ldots, \Gamma_g$ be subsets of $\{1, \ldots, K\}$ such that for $j = 1, \ldots, g$, the $j^{th}$ decoding group is $\{x_i | i \in \Gamma_j\}$. Consider a $pg2^{\lfloor \frac{g-1}{2} \rfloor} \times p2^{\lfloor \frac{g-1}{2} \rfloor}$ STBC $\mathcal{C}'$, given by $\mathcal{C}' = \{\phi(C, g) | C \in \mathcal{C}\}$. That is, $\mathcal{C}' = \{\sum_{i=1}^{K} a_i \phi(A_i, g) | [a_1, \ldots, a_K]^T \in \mathcal{A}\}$ which can be obtained from the design $\mathbf{Y} = \sum_{i=1}^{K} y_i \phi(A_i, g)$ in real symbols $y_i$, and the signal set $\mathcal{A}$.

*Proposition 6:* $\mathcal{C}'$ is a $g$-group ML-decodable, full-diversity code of rate $R' = \frac{N}{g2^{g-1}} + \frac{g^2-g}{2N}$ cspcu.

*Proof:* Since $A_1, \ldots, A_K$ are linearly independent, $\phi(A_1, g), \ldots, \phi(A_K, g)$ are also linearly independent over $\mathbb{R}$. Thus, it is clear that the rate of the code in complex symbols per channel use is $R'$.

Let $1 \leq j_1 < j_2 \leq g$ and $i_1 \in \Gamma_{j_1}$, $i_2 \in \Gamma_{j_2}$. For $l = 1, 2$, let $A_{i_l} = diag(A_{i_l}[1], \ldots, A_{i_l}[g])$. Since $x_{i_1}$ and $x_{i_2}$ belong to different decoding groups of the code $\mathcal{C}$, we have $A_{i_1}^H A_{i_2} + A_{i_2}^H A_{i_1} = \mathbf{0}$. This implies that for each $k = 1, \ldots, g$, $A_{i_1}[k]^H A_{i_2}[k] + A_{i_2}[k]^H A_{i_1}[k] = \mathbf{0}$. Hence,

$$\phi(A_{i_1}, g)^H \phi(A_{i_2}, g) + \phi(A_{i_2}, g)^H \phi(A_{i_1}, g)$$
$$= \sum_{k=1}^{g} \left( A_{i_1}[k]^H A_{i_2}[k] + A_{i_2}[k]^H A_{i_1}[k] \right)$$
$$= \mathbf{0}.$$

Hence, for $j = 1, \ldots, g$, the $j^{th}$ decoding group is $\{y_i | i \in \Gamma_j\}$. Since $\mathcal{C}$ is balanced, $\mathcal{C}'$ is also balanced. It only remains to show that the code $\mathcal{C}'$ offers full-diversity. Consider any two distinct codewords $C_1'$ and $C_2'$ in $\mathcal{C}'$. There exist distinct $C_1, C_2 \in \mathcal{C}$ such that, $C_1' = \phi(C_1, g)$ and $C_2' = \phi(C_2, g)$.

$$\Delta C' = C_1' - C_2' = \phi(C_1, g) - \phi(C_2, g) = \phi(C_1 - C_2, g).$$

For $l = 1, 2$, let $C_l = diag(C_l[1], \ldots, C_l[2])$. Let $\Delta C = C_1 - C_2 = diag(\Delta C[1], \ldots, \Delta C[g])$. Since $\mathcal{C}$ offers full-diversity, we have, $det(\Delta C) = \prod_{k=1}^{g} det(\Delta C[k]) \neq 0$. Hence, for each $k = 1, \ldots, g$, $det(\Delta C[k]) \neq 0$. Thus,

$$det(\Delta C'^H \Delta C') = det\left(\sum_{k=1}^{g} \Delta C[k]^H \Delta C[k]\right) \neq 0.$$

This follows from the fact that, the sum of two positive definite matrices is also positive definite. We have thus shown that the STBC $\mathcal{C}'$ offers full-diversity. ∎

For each $g > 1$ and $p \geq 1$, we have constructed a $g$-group ML-decodable STBC for number of transmit antennas $N' = p2^{\lfloor \frac{g-1}{2} \rfloor}$, delay $T' = gN'$ and rate $R'(N') = \frac{N'}{2^{g-1}} + \frac{g-1}{2N'}$. The asymptotic normalized rate of the family of codes corresponding to a fixed $g$ is $r(g) = \frac{1}{2^{g-1}}$. The asymptotic normalized rate of these codes is better than those of the codes in Section IV by a factor of $g$. However, the delay is larger by the same factor $g$.

*Example 13:* Consider the 3-group ML-decodable code $\mathbf{X_{12\times 12}}$ (7). Define $\mathbf{X_{12\times 4}} = \phi(\mathbf{X_{12\times 12}}, 3)$. Then, $\mathbf{X_{12\times 4}}$ is a full-diversity, rate $5/4$, 3-group ML-decodable code for 4 transmit antennas with a delay of 12 channel uses. The code is comprised of 30 real symbols $\{x_1, \ldots, x_{30}\}$. With the variables as defined in Example 11, the resulting code is

$$\mathbf{X_{12\times 4}} = \begin{bmatrix} w_1 & -u_1^* & 0 & -v_1^* \\ u_1 & w_1^* & -y_1^* & 0 \\ 0 & y_1 & w_1 & -z_1^* \\ v_1 & 0 & z_1 & w_1^* \\ w_2 & -u_2^* & 0 & -v_2^* \\ u_2 & w_2^* & -y_2^* & 0 \\ 0 & y_2 & w_2 & -z_2^* \\ v_2 & 0 & z_2 & w_2^* \\ w_3 & -u_3^* & 0 & -v_3^* \\ u_3 & w_3^* & -y_3^* & 0 \\ 0 & y_3 & w_3 & -z_3^* \\ v_3 & 0 & z_3 & w_3^* \end{bmatrix}.$$

### D. Comparison with other Multigroup ML-decodable codes

We consider only balanced multigroup ML-decodable codes available in the literature. A summary of comparison is given in Table I. The delay $T$ of maximum rate orthogonal designs is lower bounded by $\sigma(N)$, where for $m > 0$, $\sigma(2m) = \sigma(2m-1) = \binom{2m}{m-1}$ [31]. Delay-optimal 2-group ML-decodable codes in [17] have asymptotic-normalized-rate $r = 1/4$, but were constructed only for $2^m$ antennas. We have constructed 2-group ML-decodable, delay-optimal codes for all antennas $N \geq 2$ in Section III-B. For the particular cases of $N = 4$ and $N = 2^m$ these new codes have the same rates as that of the codes in [16] and [17] respectively. The 2-group ML-decodable codes of Section III-C have the same rate as that of the codes in [18] for identical delay $T = 2N$.

The new codes constructed in this section are the first instances of asymptotically-good $g$-group ML-decodable codes reported in the literature for $g > 2$.

TABLE I
COMPARISON OF BALANCED MULTIGROUP ML-DECODABLE CODES

| Code | Transmit Antennas $N$ | Delay $T$ | Groups $g$ | Rate (cspcu) $R$ | Asymptotic rate | Asymptotic normalized rate $r$ |
|---|---|---|---|---|---|---|
| Square COD[1] | $2^m, m \geq 0$ | $N$ | $2RT$ | $\frac{m+1}{2^m}$ | 0 | 0 |
| Maximal Rate COD[1] | $\geq 1$ | at least $\sigma(N)^\dagger$ | $2RT$ | $\frac{\lceil \frac{N-1}{2} \rceil + 1}{2\lceil \frac{N-1}{2} \rceil}$ | $\frac{1}{2}$ | 0 |
| SSD Codes [2] | $2^m, m \geq 1$ | $T$ | $RT$ | $\frac{m}{2^{m-1}}$ | 0 | 0 |
| CUWDs[3] [3], [14] Natarajan et. al. [15] | $2^m, m \geq \lceil \frac{g}{2} - 1 \rceil$ | $N$ | $\geq 2$ | $\frac{g}{2^{\lfloor \frac{g+1}{2} \rfloor}}$ | $\frac{g}{2^{\lfloor \frac{g+1}{2} \rfloor}}$ | 0 |
| Dao et al. [1] de Abreu [13] | $2^m, m \geq 1$ | $N$ | 4 | 1 | 1 | 0 |
| Dao et al. [1] Rajan et al. [3] | $2m, m \geq 1$ | $N$ | 4 | 1 | 1 | 0 |
| Yuen et al. [16] | 4 | 4 | 2 | $5/4$ | NA | NA |
| Srinath et al. [17] | $2^m, m \geq 2$ | $N$ | 2 | $\frac{N}{4} + \frac{1}{N}$ | $\infty$ | $\frac{1}{4}$ |
| Ren et al. [18] | $\geq 1$ | $\geq 2N$ | 2 | $\frac{NT - N^2 + 1}{T}$ | $\infty$ | $\frac{1}{2}$ for $T = 2N$ |
| **New codes in Sec III-B** | $\mathbf{m2^{\lfloor \frac{g-1}{2} \rfloor}, m \geq g}$ | $\mathbf{N}$ | $\geq 2$ | $\frac{g}{N(1+\mathbf{1}\{g \text{ is even}\})} \left\lfloor \frac{N}{g2^{\lfloor \frac{g-1}{2} \rfloor}} \right\rfloor^2 + \frac{g^2 - g}{2N}$ | $\infty$ | $\frac{1}{g2^{g-1}}$ |
| **New codes in Sec III-C** | $\mathbf{m2^{\lfloor \frac{g-1}{2} \rfloor}, m \geq 1}$ | $\mathbf{gN}$ | $\geq 2$ | $\frac{N}{2^{g-1}} + \frac{g-1}{2N}$ | $\infty$ | $\frac{1}{2^{g-1}}$ |
| Division Algebra codes [21] | $\geq 1$ | $N$ | 1 | $N$ | $\infty$ | 1 |

[†] For $m \geq 1$, $\sigma(2m) = \sigma(2m-1) = \binom{2m}{m-1}$.
Key : 1) COD - Complex Orthogonal Design, 2) SSD Codes - Single complex symbol ML-decodable codes,
3) CUWDs - Clifford Unitary Weight Designs.

## IV. A NEW CLASS OF FAST-GROUP-DECODABLE CODES

In this section, we construct a new class of FGD codes for all even number of transmit antennas and rates $1 < R \leq 5/4$. In [15], FGD codes for the same range of rates, but for a smaller class of antennas, number of antennas that are a power of 2, were constructed. The FGD codes of this section match the FGD codes of [15] in terms of ML decoding complexity when the number of antennas is a power of 2. All the codes considered in this section are delay-optimal.

### A. A new class of FGD codes

We first construct a rate $5/4$ FGD code for $2m$ antennas, $m \geq 1$. Codes of rates less than $5/4$ are then obtained via puncturing. Let the number of transmit antennas $N = 2m$, for some positive integer $m$. Let $X = \begin{bmatrix} 0 & 1 \\ 1 & 0 \end{bmatrix}$ and $Z = \begin{bmatrix} 1 & 0 \\ 0 & -1 \end{bmatrix}$. $I_2, iX, iZ$ and $ZX$ are the weight matrices of the Alamouti code and are mutually Hurwitz-Radon orthogonal. Define $\mathcal{D}_m \subset \mathbb{C}^{m \times m}$ as

$$\mathcal{D}_m = \{diag(1,1,1,\ldots,1), diag(1,-1,1,\ldots,1),$$
$$diag(1,1,-1,\ldots,1),\ldots,diag(1,1,1,\ldots,-1)\}.$$

All the matrices in $\mathcal{D}_m$ are Hermitian and mutually commuting and $|\mathcal{D}_m| = m$. Let $\mathcal{P} = \{I_2, iX, iZ, ZX, iI_2\}$. Let $\mathcal{P}_1 = \{I_2\}$ and $\mathcal{P}_2 = \{iX, iZ, ZX, iI_2\}$. For any $\mathcal{S} \subset \mathbb{C}^{2 \times 2}$, let $\mathcal{S} \otimes \mathcal{D}_m = \{A \otimes B | A \in \mathcal{S}, B \in \mathcal{D}_m\}$. The proposed $N \times N$ design is $\mathbf{X_{FGD}} = \sum_{A \in \mathcal{P} \otimes \mathcal{D}_m} x_A A$.

*Proposition 7:* The rate of the design $\mathbf{X_{FGD}}$ is $5/4$ cspcu.

*Proof:* $|\mathcal{P} \otimes \mathcal{D}_m| = |\mathcal{P}||\mathcal{D}_m| = 5m$. The rate is $\frac{5m/2}{N} = 5/4$ cspcu provided the set $\mathcal{P} \otimes \mathcal{D}_m$ is linearly independent over $\mathbb{R}$. It is therefore enough to show that both the sets $\mathcal{P}$ and $\mathcal{D}_m$ are linearly independent over $\mathbb{R}$. Proving linear independence for $\mathcal{P}$ is straightforward once the matrices are vectorized. We now show that the set $\mathcal{D}_m$ is linearly independent. Since the elements of $\mathcal{D}_m$ are diagonal this is equivalent to showing that the following $m \times m$ matrix, whose $m$ columns are the diagonal entries of the $m$ matrices in $\mathcal{D}_m$, is full-rank,

$$\begin{bmatrix} 1 & 1 & 1 & \cdots & 1 \\ 1 & -1 & 1 & \cdots & 1 \\ 1 & 1 & -1 & \cdots & 1 \\ \vdots & \vdots & \vdots & \ddots & \vdots \\ 1 & 1 & 1 & \cdots & -1 \end{bmatrix} = \mathbf{1}\mathbf{1}^T - diag(0,2,\ldots,2), \quad (9)$$

where $\mathbf{1}$ is the all ones vector. Let $a = [a_1, \ldots, a_m]^T \in \mathbb{R}^m$ be in the null-space of (9). Then we have,

$$\mathbf{1}\left(\sum_{n=1}^{m} a_n\right) = diag(0, 2a_2, \ldots, 2a_m).$$

Equating the components of the vectors on either side of the equation, we have $0 = \sum_{n=1}^{m} a_n = 2a_k$ for $k > 1$. Thus, $0 = \sum_{n=1}^{m} a_n = a_1$. This completes the proof. ∎

We now show that the proposed code is fast-group-decodable. Let $\mathbf{X_1} = \sum_{A \in \mathcal{P}_1 \otimes \mathcal{D}_m} x_A A$ and $\mathbf{X_2} = \sum_{A \in \mathcal{P}_2 \otimes \mathcal{D}_m} x_A A$. We have that $\mathbf{X_{FGD}} = \mathbf{X_1} + \mathbf{X_2}$.

*Proposition 8:* The design $\mathbf{X_{FGD}}$ is 2-group ML decodable. The two groups correspond to the symbols in the designs $\mathbf{X_1}$ and $\mathbf{X_2}$ respectively.

*Proof:* It is enough to show that for any $A \in \mathcal{P}_1 \otimes \mathcal{D}_m$ and any $B \in \mathcal{P}_2 \otimes \mathcal{D}_m$, we have $A^H B + B^H A = \mathbf{0}$. Equivalently, $A^H B$ must be skew-Hermitian. $A$ and $B$ are of the form $A = I_2 \otimes D_A$ and $B = S \otimes D_B$, where $S \in \mathcal{P}_2$ is skew-Hermitian, $D_A$ and $D_B$ are diagonal, commuting and Hermitian. Thus, $A^H B = S \otimes D_A D_B$, which is a skew-Hermitian matrix. This completes the proof. ∎

*Proposition 9:* The design $\mathbf{X_2}$ is conditionally 3-group ML decodable i.e. fast-decodable.

*Proof:* It is enough to show that the design $\sum_{A \in \{iX, iZ, ZX\} \otimes \mathcal{D}_m} x_A A$, whose weight matrices form a subset of weight matrices of $\mathbf{X_2}$, is 3 group ML decodable. It is straightforward to show the matrices belonging to the three sets $\{iX\} \otimes \mathcal{D}_m$, $\{iZ\} \otimes \mathcal{D}_m$ and $\{ZX\} \otimes \mathcal{D}_m$ are mutually Hurwitz-Radon orthogonal. ∎

From Propositions 8 and 9, it is clear that $\mathbf{X_{FGD}}$ is 2-group ML decodable and one of the groups is fast-decodable, proving our claim that $\mathbf{X_{FGD}}$ is fast-group-decodable. The rate-1, 4-group ML decodable design given in [2] has the elements of $\{I_2, iX, iZ, ZX\} \otimes \mathcal{D}_m$ as weight matrices. The new design $\mathbf{X_{FGD}}$ contains the matrices in $\{iI_2\} \otimes \mathcal{D}_m$ in addition to the weight matrices of the design in [2].

Designs of rate $1 < R \leq 5/4$ can be obtained by puncturing appropriate number of symbols from the set of symbols corresponding to $\{iI_2\} \otimes \mathcal{D}_m$. Similar to Propositions 8 and 9, it can be shown that the designs thus obtained are fast-group-decodable. The resulting design will be 2-group ML decodable, the first group containing symbols corresponding to $\mathcal{P}_1 \otimes \mathcal{D}_m$ and the second group containing the rest of the symbols. Again, the second group is conditionally 3-group ML decodable, the three conditional groups correspond to the three sets $\{iX\} \otimes \mathcal{D}_m$, $\{iZ\} \otimes \mathcal{D}_m$ and $\{ZX\} \otimes \mathcal{D}_m$.

### B. ML Decoding Complexity of the new class of FGD codes

We assume encoding using full-diversity achieving, regular-PAM constellation as promised by Corollary 2. Let the size of the underlying complex constellation used for encoding the proposed design of rate $R$, $1 < R \leq 5/4$, be $M$. Let the number of transmit antennas be $N = 2m$. The group of symbols corresponding to $\mathcal{P}_1 \otimes \mathcal{D}_m$ can be ML decoded independently of other symbols. For each of the $M^{\frac{m-1}{2}}$ values that a subset of $m-1$ real symbols of $\mathcal{P}_1 \otimes \mathcal{D}_m$ can take, the conditionally optimal value of the last real symbol can be found by simple scaling and hard limiting since its constellation is a regular PAM. The first group can thus be ML decoded with complexity $M^{\frac{m-1}{2}}$. The second group consists of symbols that correspond to $\{iX\} \otimes \mathcal{D}_m$, $\{iZ\} \otimes \mathcal{D}_m$, $\{ZX\} \otimes \mathcal{D}_m$ and a subset $\mathfrak{D} \subset \{iI_2\} \otimes \mathcal{D}_m$ of cardinality $4m(R-1)$. For each of the $M^{2m(R-1)}$ values that the symbols of $\mathfrak{D}$ can jointly take, the three groups of symbols corresponding to $\{iX\} \otimes \mathcal{D}_m$, $\{iZ\} \otimes \mathcal{D}_m$ and $\{ZX\} \otimes \mathcal{D}_m$ can be independently decoded, each with complexity $M^{\frac{m-1}{2}}$. Hence, the proposed code can be ML decoded with a complexity of $M^{m/2-0.5} + M^{2m(R-1)} \cdot 3M^{m/2-0.5}$. For large $M$, this is approximately equal to

$$3M^{\frac{N}{4}(4R-3)-0.5}. \qquad (10)$$

The FGD codes of [15], constructed for number of antennas $N$ that are a power of 2 and rates $1 < R \leq 5/4$, can be ML decoded with complexity $3M^{\frac{N}{4}(4R-3)-0.5}$. Thus, the ML decoding complexity of the new FGD codes of this section match the ML decoding complexity of FGD codes from [15] when the number of antennas is a power of 2.

## V. A NEW CLASS OF ASYMPTOTICALLY-OPTIMAL, FAST-DECODABLE, FULL-DIVERSITY STBCs

In this section, we first give a general procedure to construct high-rate, full-diversity FD codes using lower rate multigroup ML decodable codes and FGD codes. We then construct new STBCs for rates $R > 1$ and number of antennas $N > 1$ using the delay-optimal, asymptotically-good multigroup ML decodable codes of Section III-B and the FGD codes of Section IV. Finally we show that the new class of codes have the least ML decoding complexity among all the codes available in the literature for $N > 1$ and $R > 1$. Among the new codes, the class of full-rate codes are asymptotically-optimal and fast-decodable, and for $N > 5$, have lower ML decoding complexity than all other known families of asymptotically-optimal, full-diversity, FD codes. All the codes considered in this section are delay-optimal.

### A. A new class of low ML decoding complexity STBCs from multigroup ML decodable codes and FGD codes

Let $\mathbf{X_b} = \sum_{i=1}^{K_b} x_i A_i$ be an $N \times N$ design of rate $R_b = K_b/2N$ cspcu. We have the following propositions.

*Proposition 10:* Let $R_b < R \leq N$ and $K = 2RN$ and $\mathcal{B}$ be an $\mathbb{R}$-basis of $\mathbb{C}^{N \times N}$. Then, there exist $K - K_b$ matrices $A_{K_b+1}, \ldots, A_K \in \mathcal{B}$, such that the set $\{A_1, \ldots, A_K\}$ is linearly independent over $\mathbb{R}$.

*Proof:* Let $\mathcal{B} = \{U_1, \ldots, U_{2N^2}\}$. Assign $C_i = A_i$, $i = 1, \ldots, K_b$ and $C_i = U_{i-K_b}$ for $i = K_b+1, \ldots, K_b+2N^2$. Consider the set $\mathcal{S}_0 = \{1, \ldots, K+2N^2\}$. For any $\mathcal{S} \subset \mathcal{S}_0$, define $\mathcal{M}(\mathcal{S}) = \{C_i | i \in \mathcal{S}\}$. Since $\mathcal{B}$ is a $\mathbb{R}$-linear basis for $\mathbb{C}^{N \times N}$, $span_{\mathbb{R}}(\mathcal{M}(\mathcal{S}_0)) = \mathbb{C}^{N \times N}$.

Find the smallest $j \in \mathcal{S}_0$ such that $C_j \in span_{\mathbb{R}}(\{C_1, \ldots, C_{j-1}\})$. Such a $j$ always exists because the set $\mathcal{M}(\mathcal{S}_0)$ is linearly dependent. Since $\mathbf{X_b}$ is a design, its weight matrices are linearly independent over $\mathbb{R}$. Hence, $C_1, \ldots, C_{K_b}$ are linearly independent over $\mathbb{R}$ and thus we have $j > K_b$. Let $\mathcal{S}_1 = \mathcal{S}_0 \setminus \{j\}$. Thus, $\{1, \ldots, K_b\} \subset \mathcal{S}_1$ and $span_{\mathbb{R}}(\mathcal{M}(\mathcal{S}_1)) = \mathbb{C}^{N \times N}$. In general for $i \geq 1$, if $\mathcal{M}(\mathcal{S}_{i-1})$ is linearly dependent, find the smallest $j \in \mathcal{S}_{i-1}$ such that $C_j \in span_{\mathbb{R}}(\{k \in \mathcal{S}_{i-1} | k < j\})$ and let $\mathcal{S}_i = \mathcal{S}_{i-1} \setminus \{j\}$. It is clear that, $span_{\mathbb{R}}(\mathcal{M}(\mathcal{S}_i)) = \mathbb{C}^{N \times N}$ and $\{1, \ldots, K_b\} \subset \mathcal{S}_i$. Since $\mathcal{S}_0$ is a finite set, there exists a finite $l$ such that $\mathcal{M}(\mathcal{S}_l)$ is linearly independent over $\mathbb{R}$ while $\mathcal{M}(\mathcal{S}_{l-1})$ is not. Also, by construction, $span_{\mathbb{R}}(\mathcal{M}(\mathcal{S}_l)) = \mathbb{C}^{N \times N}$. Thus, $\mathcal{M}(\mathcal{S}_l)$ is a basis for $\mathbb{C}^{N \times N}$ and $\mathcal{M}(\mathcal{S}_l)$ contains $A_1, \ldots, A_{K_b}$.

From the hypothesis of the proposition, $K \leq 2N^2$ and $K - K_b \leq 2N^2 - K_b$. Label any of the $K - K_b$ matrices from $\mathcal{M}(\mathcal{S}_l) \setminus \{A_1, \ldots, A_{K_b}\}$ as $A_{K_b+1}, \ldots, A_K$. This completes the proof. ∎

*Proposition 11:* Let $k \geq 1$. There exists an explicitly constructable $\mathbb{R}$-basis $\mathcal{B}_k$ of $\mathbb{C}^{k \times k}$, consisting of unitary matrices.

*Proof:* Let $\mathcal{H}_k$ be the explicitly constructable set of $k^2$ unitary, Hermitian, $\mathbb{R}$-linearly independent $k \times k$ complex matrices from Proposition 4. Let $\mathcal{B}_k = \mathcal{H}_k \cup i\mathcal{H}_k$. Let $span_{\mathbb{R}}(\mathcal{H}_k)$ denote the subspace spanned by elements of $\mathcal{H}_k$ over $\mathbb{R}$. $\mathcal{B}_k$ is explicitly constructable since $\mathcal{H}_k$ is explicitly constructable. Also, $|\mathcal{B}_k| = 2k^2$. Since the dimension of $\mathbb{C}^{k \times k}$ as a vector space $\mathbb{R}$ is $2k^2$, it is enough to show that $dim_{\mathbb{R}}(span_{\mathbb{R}}(\mathcal{B}_k)) = 2k^2$, where $dim_{\mathbb{R}}(\cdot)$ denotes the dimension of a subspace over $\mathbb{R}$. We have that,

$$dim_{\mathbb{R}}(span_{\mathbb{R}}(\mathcal{B}_k))$$
$$= dim_{\mathbb{R}}(span_{\mathbb{R}}(\mathcal{H}_k)) + dim_{\mathbb{R}}(span_{\mathbb{R}}(i\mathcal{H}_k))$$
$$\quad - dim_{\mathbb{R}}(span_{\mathbb{R}}(\mathcal{H}_k) \cap span_{\mathbb{R}}(i\mathcal{H}_k))$$
$$= 2k^2 - dim_{\mathbb{R}}(span_{\mathbb{R}}(\mathcal{H}_k) \cap span_{\mathbb{R}}(i\mathcal{H}_k)).$$

Thus, it is enough to show that $dim_{\mathbb{R}}(span_{\mathbb{R}}(\mathcal{H}_k) \cap span_{\mathbb{R}}(i\mathcal{H}_k)) = 0$. Let $A$ belong to the intersection of the two subspaces $span_{\mathbb{R}}(\mathcal{H}_k)$ and $span_{\mathbb{R}}(i\mathcal{H}_k)$. This would mean that $A$ is both Hermitian and skew-Hermitian. Thus, $A = A^H = -A$. Hence, $A = \mathbf{0}$. This completes the proof. ∎

The result of Proposition 10 together with the $\mathbb{R}$-basis $\mathcal{B}_k$ promised by Proposition 11 shows that any design $\mathbf{X_b}$, with rate $R_b < N$, can be extended by adding sufficient number of real symbols with full-rank weight matrices. The new design from Proposition 10 is $\mathbf{X} = \sum_{i=1}^{K} x_i A_i$. We say that $\mathbf{X_b}$ is the *base design* of $\mathbf{X}$.

Consider the case when all the weight matrices of the base design $\mathbf{X_b}$ are full-rank. Then, all the weight matrices of the design $\mathbf{X}$ are also full-ranked. For any given $Q$, from Corollary 2, there exist $Q$-ary regular PAM constellations $\mathcal{A}_{d_1,Q}, \ldots, \mathcal{A}_{d_K,Q}$ with minimum Euclidean distance $d_1, \ldots, d_K$ respectively, such that the STBC $\mathcal{C}(\mathbf{X}, \mathcal{A}_{d_1,Q} \times \cdots \times \mathcal{A}_{d_K,Q})$ is of full-diversity. Let $M$ be the size of the underlying complex constellation used, i.e., $M = Q^2$. ML decoding can be performed by finding the optimal values of $x_1, \ldots, x_{K_b}$ for each of the $M^{(K-K_b)/2}$ values that $x_{K_b+1}, \ldots, x_K$ jointly assume and then by choosing the tuple $x_1, \ldots, x_K$ that optimizes the ML metric among the $M^{(K-K_b)/2}$ tuples from the first step. The problem of optimizing $x_1, \ldots, x_{K_b}$ for a given set of values for $x_{K_b+1}, \ldots, x_K$ is equivalent to the problem of decoding the STBC $\mathcal{C}(\mathbf{X_b}, \mathcal{A}_{d_1,Q} \times \cdots \times \mathcal{A}_{d_{K_b},Q})$. Thus the STBC $\mathcal{C}(\mathbf{X}, \mathcal{A}_{d_1,Q} \times \cdots \times \mathcal{A}_{d_K,Q})$ can be ML decoded with a complexity given in (11) at the top of the next page, where $ML\text{-}decoding\text{-}complexity(\cdot)$ denotes the ML decoding complexity of an STBC. Further if $\mathcal{C}(\mathbf{X_b}, \mathcal{A}_{d_1,Q} \times \cdots \times \mathcal{A}_{d_{K_b},Q})$ were multigroup ML decodable or fast-group-decodable, the proposed code $\mathcal{C}(\mathbf{X}, \mathcal{A}_{d_1,Q} \times \cdots \times \mathcal{A}_{d_K,Q})$ will be fast-decodable.

*Example 14:* Consider the FD code for $N = 4$ and $R = \frac{3}{2}$ cspcu constructed in [28]. There are 12 real symbols in this design $\mathbf{X} = \sum_{i=1}^{12} x_i A_i$. The base design $\mathbf{X_b} = \sum_{i=1}^{6} x_i A_i$ is the maximal rate $4 \times 4$ complex orthogonal design [8]. The matrices $A_7, \ldots, A_{12}$ are chosen as $A_{i+6} = A_i \Lambda'$ for $i = 1, \ldots, 6$, where $\Lambda' \in \mathbb{C}^{4 \times 4}$ is a full-rank matrix chosen to maximize the diversity and coding gain. The real symbols are encoded independently from a regular PAM constellation of unit minimum Euclidean distance, $\mathcal{A}_{Q-PAM}$. Since $\mathbf{X_b}$ is an orthogonal design, $\mathcal{C}(\mathbf{X_b}, \mathcal{A}_{Q-PAM} \times \cdots \times \mathcal{A}_{Q-PAM})$ can be ML decoded with $O(1)$ complexity. Hence, from (11), the FD STBC $\mathcal{C}(\mathbf{X}, \mathcal{A}_{Q-PAM} \times \cdots \times \mathcal{A}_{Q-PAM})$ can be ML decoded with a complexity of the order of $M^3$.

From the above discussion, it is clear that new FD codes of rates $R > R_b$ can be constructed using already known multigroup ML-decodable and FGD codes of rate $R_b$. When we are interested in rates $R < R_b$, low ML decoding complexity codes can be obtained by puncturing known multigroup ML decodable and FGD codes of rate $R_b$. In this case too, we say that $\mathbf{X_b}$ is the base design of the new code. It is straightforward to show that the order of ML decoding complexity, when the base design is $g$-group ML decodable and $R < R_b$, is

$$M^{(\lceil \frac{K}{g} \rceil - 1)/2}. \tag{12}$$

### B. Explicit constructions and their ML decoding complexity

In this subsection, we explicitly construct STBCs with low ML decoding complexity (multigroup ML decodable, FD and FGD) for rates greater than 1 cspcu and $N > 1$ transmit antennas. To do so we start from base designs of rate 1 cspcu or higher. From (11) it is clear that for the constructed code to be of low ML decoding complexity, the base design itself should be of low ML decoding complexity. To the best of the authors' knowledge, multigroup ML decodable and FGD codes from among the following three classes of codes offer the least known ML decoding complexity for rates $R \geq 1$ and number of antennas $N \geq 1$.

- $\mathcal{F}_{DAST}$: 2-group ML decodable, full-diversity, Diagonal Algebraic Space-Time (DAST) block codes from [32] for any number of transmit antennas $N$ and rate $R = 1$ cspcu.
- $\mathcal{F}_{g,AG}$, $g > 1$: $g$-group ML decodable, delay-optimal, asymptotically-good STBCs in Section III-B of this paper for number of antennas, $N = n2^{\lfloor \frac{g-1}{2} \rfloor}$, $n \geq g$.

$$M^{(K-K_b)/2} \times ML\text{-}decoding\text{-}complexity\left(\mathcal{C}(\mathbf{X_b}, \mathcal{A}_{d_1,Q} \times \cdots \times \mathcal{A}_{d_{K_b},Q})\right) \quad (11)$$

TABLE II
EXPONENT OF $M^\dagger$ IN THE ORDER OF ML DECODING COMPLEXITY OF THE NEW STBCS AND THEIR ASSOCIATED BASE DESIGNS

| Transmit Antennas $N$ | Rate, $R$ in cspcu | | | | | | | | | |
|---|---|---|---|---|---|---|---|---|---|---|
| | 5/4 | 2 | 3 | 4 | 5 | 6 | 7 | 8 | 9 | 10 |
| 2 | 0.5 $\mathcal{F}_{FGD}$ | 2 $\mathcal{F}_{FGD}$ | | | | | | | | |
| 3 | 2 $\mathcal{F}_{DAST}$ | 4 $\mathcal{F}_{DAST}$ | 7 $\mathcal{F}_{DAST}$ | | | | | | | |
| 4 | 1.5 $\mathcal{F}_{FGD}$ | 4.5 $\mathcal{F}_{FGD}$ | 8.5 $\mathcal{F}_{FGD}$ | 12.5 $\mathcal{F}_{FGD}$ | | | | | | |
| 5 | 3.5 $\mathcal{F}_{DAST}$ | 7 $\mathcal{F}_{DAST}$ | 12 $\mathcal{F}_{DAST}$ | 17 $\mathcal{F}_{DAST}$ | 22 $\mathcal{F}_{DAST}$ | | | | | |
| 6 | 2.5 $\mathcal{F}_{FGD}$ | 6.5 $\mathcal{F}_{2,AG}$ | 12.5 $\mathcal{F}_{2,AG}$ | 18.5 $\mathcal{F}_{2,AG}$ | 24.5 $\mathcal{F}_{2,AG}$ | 30.5 $\mathcal{F}_{2,AG}$ | | | | |
| 7 | 4 $\mathcal{F}_{2,AG}$ | 8.5 $\mathcal{F}_{2,AG}$ | 15.5 $\mathcal{F}_{2,AG}$ | 22.5 $\mathcal{F}_{2,AG}$ | 29.5 $\mathcal{F}_{2,AG}$ | 36.5 $\mathcal{F}_{2,AG}$ | 43.5 $\mathcal{F}_{2,AG}$ | | | |
| 8 | 3.5 $\mathcal{F}_{FGD}$ | 7.5 $\mathcal{F}_{2,AG}$ | 15 $\mathcal{F}_{2,AG}$ | 23 $\mathcal{F}_{2,AG}$ | 31 $\mathcal{F}_{2,AG}$ | 39 $\mathcal{F}_{2,AG}$ | 47 $\mathcal{F}_{2,AG}$ | 55 $\mathcal{F}_{2,AG}$ | | |
| 9 | 5.5 $\mathcal{F}_{2,AG}$ | 9 $\mathcal{F}_{2,AG}$ | 18 $\mathcal{F}_{2,AG}$ | 27 $\mathcal{F}_{2,AG}$ | 36 $\mathcal{F}_{2,AG}$ | 45 $\mathcal{F}_{2,AG}$ | 54 $\mathcal{F}_{2,AG}$ | 63 $\mathcal{F}_{2,AG}$ | 72 $\mathcal{F}_{2,AG}$ | |
| 10 | 4.5 $\mathcal{F}_{FGD}$ | 9.5 $\mathcal{F}_{2,AG}$ | 16.5 $\mathcal{F}_{2,AG}$ | 26.5 $\mathcal{F}_{2,AG}$ | 36.5 $\mathcal{F}_{2,AG}$ | 46.5 $\mathcal{F}_{2,AG}$ | 56.5 $\mathcal{F}_{2,AG}$ | 66.5 $\mathcal{F}_{2,AG}$ | 76.5 $\mathcal{F}_{2,AG}$ | 86.5 $\mathcal{F}_{2,AG}$ |

$^\dagger$ $M$ is the size of the underlying complex constellation.

- $\mathcal{F}_{FGD}$: FGD codes for even values of $N$ and rate $5/4$ cspcu in Section IV of this paper.

The codes in $\mathcal{F}_{FGD}$ and $\mathcal{F}_{g,AG}$, $g > 1$, have full-rank weight matrices. We now show that the codes in $\mathcal{F}_{DAST}$ can be obtained by using designs with full-rank weight matrices and by independent encoding of real symbols using $\mathcal{A}_{Q-PAM}$.

*Proposition 12:* Let $\mathcal{C}_0 \in \mathcal{F}_{DAST}$ be a code for $N$ transmit antennas that uses an underlying complex constellation of size $Q^2$. There exists an $N \times N$ design $\mathbf{X_0}$ with full-rank weight matrices such that, $\mathcal{C}_0 = \mathcal{C}(\mathbf{X_0}, \mathcal{A}_{Q-PAM} \times \cdots \times \mathcal{A}_{Q-PAM})$.

*Proof:* In [32], the full-diversity STBC $\mathcal{C}_0$ was obtained by using the design $\tilde{\mathbf{X}} = diag(\tilde{x}_1 + i\tilde{x}_{N+1}, \tilde{x}_2 + i\tilde{x}_{N+2}, \ldots, \tilde{x}_N + i\tilde{x}_{2N})$. The variables $[\tilde{x}_1, \ldots \tilde{x}_N]^T$ and $[\tilde{x}_{N+1}, \ldots \tilde{x}_{2N}]^T$ are encoded independently using the $N$ dimensional signal set $\mathcal{A} = \{U[a_1, \ldots, a_N]^T | a_i \in \mathcal{A}_{Q-PAM}, i = 1, \ldots, N\}$, where $U \in \mathbb{R}^{N \times N}$ is an orthogonal matrix. Let $U = [u_1 u_2 \cdots u_N]$ and $[x_{1+i}, \ldots, x_{N+i}]^T = U^{-1}[\tilde{x}_{1+i}, \ldots, \tilde{x}_{N+i}]^T$ for $i = 0, N$. Further, let $A_k = diag(u_k)$ for $k = 1, \ldots, N$ and $A_k = i \cdot diag(u_{k-N})$ for $k = N+1, \ldots, 2N$. It is straightforward to show that $\mathcal{C}_0 = \mathcal{C}(\mathbf{X_0}, \mathcal{A}_{Q-PAM} \times \cdots \times \mathcal{A}_{Q-PAM})$, where $\mathbf{X_0} = \sum_{i=1}^{2N} x_i A_i$. Since $\mathcal{C}_0$ has full diversity, we have that $\zeta A_k$, $k = 1, \ldots, 2N$, is full rank for $\zeta \in \Delta \mathcal{A}_{Q-PAM} \setminus \{0\}$. Thus, all the weight matrices of the design $\mathbf{X_0}$ have full-rank. ∎

Suppose we are interested in constructing a low ML decoding complexity STBC for number of antennas $N$ and rate $R$ cspcu, where $R > 1$. The base design for this code are to be chosen from the designs in $\mathcal{F}_{FGD}$, $\mathcal{F}_{DAST}$ and $\mathcal{F}_{g,AG}$, $g > 1$, that are of size $N \times N$. The set of such designs is finite and non-empty. It is a non-empty set since $\mathcal{F}_{DAST}$ has one $N \times N$ design for each $N \geq 1$. Now, for each of the designs in this set, we compute the complexity of ML decoding the code of rate $R$ that uses this design as the base. This can be computed using (5) (10), (11) and (12). Then, from among these designs, we pick that design as the base which leads to a rate $R$ code of least ML decoding complexity.

Table II gives the achievable order of ML decoding complexities and the associated base design for a set of rates $R$ and number of antennas $N = 2, \ldots, 10$. These values were found through a computer search. A computer search for this problem is easy to implement and executes quickly. For $N = 2$ and 4 antennas, using the new FGD design of Section IV as the base gives the least known decoding complexity for any rate $1 < R \leq N$ and not just the rates discussed in the table. For $N = 3, 5$ and $N = 7, 9$, the DAST codes and the 2-group ML decodable codes of Section III-B, respectively, lead to the least known ML decoding complexity for rates $1 < R \leq N$. For $N = 6, 8, 10$, one should use the new FGD designs from Section IV for $1 < R < 3/2$ and the 2-group ML decodable designs of Section III-B for $3/2 \leq R \leq N$.

*Example 15:* We now give a rate, $R = 2$, $N = 6$ antenna design with full-rank weight matrices. When the real symbols are encoded independently using regular PAM constellations, the order of ML decoding complexity of the resulting STBC is $M^{6.5}$. The order of ML decoding complexity of codes

$$\mathbf{X_{FD}} = \begin{bmatrix} z_8 + iz_1 & -x_5 + ix_4 & -x_7 + ix_6 & x_{21} + ix_{22} & x_{23} + ix_{24} & 0 \\ x_5 + ix_4 & z_8 + iz_2 & -x_9 + ix_8 & 0 & x_{21} + ix_{22} & x_{23} + ix_{24} \\ x_7 + ix_6 & x_9 + ix_8 & z_8 + iz_3 & x_{23} + ix_{24} & 0 & x_{21} + ix_{22} \\ x_{21} + ix_{22} & x_{23} + ix_{24} & 0 & z_7 + iz_4 & -x_{15} + ix_{14} & -x_{17} + ix_{16} \\ 0 & x_{21} + ix_{22} & x_{23} + ix_{24} & x_{15} + ix_{14} & z_7 + iz_5 & -x_{19} + ix_{18} \\ x_{23} + ix_{24} & 0 & x_{21} + ix_{22} & x_{17} + ix_{16} & x_{19} + ix_{18} & z_7 + iz_6 \end{bmatrix} \quad (13)$$

in [20] and [27] for these parameters are $M^{8.5}$ and $M^8$ respectively (see Table III). For the new code, number of real symbols $K = 24$, $K_b = 20$ and the base design is the rate-5/3, 2-group ML decodable code $\mathbf{X_{6 \times 6}}$ (6) constructed Section III-B (see Example 10). Denote the rate-2, FD design by $\mathbf{X_{FD}} = \sum_{i=1}^{24} x_i A_i$ and let the base design be $\mathbf{X_{6 \times 6}} = \sum_{i=1}^{20} x_i A_i$. The design $\mathbf{X_{FD}}$ is given by (13) at the top of this page, where the auxiliary variables in the design, $z_j$, $j = 1, \ldots, 8$, are as given in Example 10.

### C. Comparison of ML decoding complexities

We now compare the ML decoding complexities of the STBCs constructed in this paper with other low ML decoding complexity codes available in the literature. In this sub-section we show that the ML decoding complexities reported in this paper are the least known in the literature.

*1) Comparison with codes from [15]:* In [15], full-diversity, FD codes were constructed for $N = 2^m$, $m \geq 1$, antennas and rates $R > 1$ and these codes use, as base designs, a family of rate-5/4 FGD codes that have the same known ML decoding complexity as the codes from $\mathcal{F}_{FGD}$. For $N = 2, 4$, and $R > 1$, the codes in [15] have least known ML decoding complexity in the literature and the new codes in Section V-B use codes from $\mathcal{F}_{FGD}$ as base for all $R > 1$. Hence, for $N = 2, 4$ and $1 < R \leq N$, the new codes match the least known ML decoding complexity in the literature.

The new codes of Section V-B either match or have lower ML decoding complexity than the codes in [15] for $N \geq 8$. The fact that the new codes match the complexities is clear, since one family of base designs $\mathcal{F}_{FGD}$ has same ML decoding complexity as the base designs of codes in [15]. We now show that the new codes have strictly lower ML decoding complexity for all $R > 3/2$ and all $N \geq 8$ that is a power of 2.

*Proposition 13:* Let $N = 2^m$, $m \geq 3$, and $3/2 < R \leq N$. The new codes for $N$ antennas and rate $R$ of Section V-B have lower ML decoding complexity than the corresponding codes of [15].

*Proof:* It is enough to show that a rate $R$ code that uses the design from $\mathcal{F}_{2,AG}$ as its base has strictly lower ML decoding complexity. We prove the result when $3/2 < R \leq \frac{N}{4} + \frac{1}{N}$. The proof for $R > \frac{N}{4} + \frac{1}{N}$ is similar and hence is omitted. Complexity of ML decoding the code that uses the design from $\mathcal{F}_{2,AG}$ as its base can be shown to be $M^{\frac{NR}{2} - \frac{1}{2}}$ for $R \leq \frac{N}{4} + \frac{1}{N}$. Complexity of the code from [15] is $M^{\frac{N}{4}(4R-3) - \frac{1}{2}}$. Hence, it is enough to show that $\frac{N}{4}(4R - 3) > \frac{NR}{2}$. But this is indeed the case whenever $R > 3/2$ cspcu. This completes the proof. ∎

The new codes of Section V-B also give lower ML decoding complexity for any $1 < R \leq 3/2$ when compared with codes from [15], but only for sufficiently large values of $N$.

*Proposition 14:* Let $R \in (1, 3/2]$. There exists a positive integer $N'$ such that for all $N > N'$ that belong to the set $\{2^m | m \in \mathbb{Z}, m > 0\}$, the new codes of Section V-B of rate $R$ and number of antennas $N$ have lower ML decoding complexity than the corresponding codes of [15].

*Proof:* It is enough to show that those codes that use designs in $\mathcal{F}_{5,AG}$ as base designs satisfy the result of the proposition. Let $g = 5$ and choose an $N' \in \{2^{\lfloor \frac{g-1}{2} \rfloor + \ell} | \ell > \lceil log_2(g) \rceil \}$ large enough such that: (1) $N' > 20/R$, and (2) for all values of $N \in \{2^{\lfloor \frac{g-1}{2} \rfloor + \ell} | \ell > \lceil log_2(g) \rceil \}$ that are greater than $N'$, the rate of the $g$-group ML decodable code from $\mathcal{F}_{g,AG}$ of size $N \times N$ is greater than $3/2$. There exists such an $N'$ since the class of codes $\mathcal{F}_{g,AG}$ is asymptotically-good, and hence their rate tends to infinity as the number of transmit antennas goes to infinity. Now choose any $N$, a power of 2, which is greater than $N'$. Consider the rate $R$ code obtained by using the $N \times N$ design from $\mathcal{F}_{g,AG}$ as the base. The rate of the base design $R_b > 3/2$ and hence the new code is obtained via puncturing and will have an ML decoding complexity of $M^{\lceil \frac{K}{2g} \rceil - \frac{1}{2}}$, where $K = 2RN$. This complexity can be upper bounded by $M^{\frac{K}{2g} + \frac{1}{2}}$. Since $N > N' > 20/R$, we have $K = 2RN > 2RN' > 40$. The codes of rate $R$ for $N$ antennas from [15] can be ML decoded with complexity $M^{\frac{N}{4}(4R-3) - \frac{1}{2}}$. Since $R > 1$, we have, $\frac{N}{4}(4R - 3) = \frac{K}{2} - \frac{3K}{8R} > \frac{K}{2} - \frac{3K}{8} = \frac{K}{8}$. Hence, it is enough to show that $\frac{K}{2g} + \frac{1}{2} < \frac{K}{8} - \frac{1}{2}$. Since, $g = 5$, this is equivalent to the condition that $K > 40$, which we have already shown to be true. This completes the proof. ∎

The value of $N$ beyond which the new codes have lower ML decoding complexity for a given $R \in (1, 3/2]$ is difficult to derive analytically, but can be found out easily through a computer search. For example, when $R = 5/4$ cspcu, the new codes of Section V-B have lower ML decoding complexity, compared with the codes from [15], for all $N \geq 32$.

*2) Comparison with TAST codes [20]:*

In [20], Threaded Algebraic Space-Time (TAST) codes were constructed for all number of transmit antennas $N \geq 1$ and rate $R = N$. These codes can be constructed by using the DAST codes as the base design. Hence, the TAST codes are conditionally 2-group ML decodable, i.e., fast-decodable. This low ML decoding complexity property of the TAST codes was not reported in [20].

We see that the new codes of Section V-B and the TAST codes have same decoding complexities for $N = 3, 5$ and any $R > 1$, since both use the DAST code as their base (see

TABLE III
COMPARISON OF ML DECODING COMPLEXITIES: EXPONENT OF $M^\dagger$ IN THE ORDER OF ML DECODING COMPLEXITY

| Transmit Antennas $N$ | Rate $R$ | **New codes** | FD[‡], FGD[‡] codes in [15] | EAST[‡] codes from [27] | TAST[‡] codes [20] | FD codes in [22] | FGD codes in [5] | FD code in [28] | FD code in [29] | Golden code [24] | Silver code [25], [26] |
|---|---|---|---|---|---|---|---|---|---|---|---|
| 2 | 1 | **0** | 0 | 0.5 | 0.5 | | | | | | |
| 2 | 2 | **2** | 2 | | 2.5 | 2.5 | | | | 2.5 | 2 |
| 4 | 1 | **0.5** | 0.5 | 1 | 1.5 | | | | | | |
| 4 | 3/2 | **2.5** | 2.5 | | 3.5 | | | 3 | | | |
| 4 | 2 | **4.5** | 4.5 | 5 | 5.5 | 4.5 | 5.5 | | 6 | | |
| 4 | 17/8 | **5** | 5 | | 6 | | 6 | | | | |
| 6 | 1 | **1** | | 1.5 | 2.5 | | | | | | |
| 6 | 2 | **6.5** | | 8 | 8.5 | | | | | | |
| 6 | 3 | **12.5** | | 14 | 14.5 | | | | | | |
| 7 | 1 | **3** | | | 3 | | | | | | |
| 7 | 2 | **8.5** | | | 10 | | | | | | |
| 7 | 3 | **15.5** | | | 17.5 | | | | | | |
| 8 | 1 | **1.5** | 1.5 | 2 | 3.5 | | | | | | |
| 8 | 2 | **7.5** | 9.5 | 10 | 11.5 | | | | | | |
| 8 | 3 | **15** | 17.5 | 18 | 19.5 | | | | | | |
| 8 | 4 | **23** | 25.5 | 26 | 27.5 | | | | | | |
| 9 | 1 | **4** | | | 4 | | | | | | |
| 9 | 2 | **9** | | | 13 | | | | | | |
| 9 | 3 | **18** | | | 22 | | | | | | |
| 9 | 4 | **27** | | | 31 | | | | | | |

[†] $M$ is the size of the underlying complex constellation.
[‡] Key: (1) FD: Fast-decodable, (2) FGD: Fast-group-decodable,
       (3) EAST: Embedded Alamouti Space-Time, (4) TAST: Threaded Algebraic Space-Time.

Table II). We now show that for any $N \geq 6$ and $R > 1$ the new codes have lower ML decoding complexity than TAST codes.

*Proposition 15:* The new codes of Section V-B for $N \geq 6$ and rates $R > 1$, have lower ML decoding complexity than the corresponding TAST codes.

*Proof:* For the TAST code of size $N \times N$, the base design is the $N \times N$ DAST code with parameters $K_b = 2N$ and $g = 2$. Also, it is clear that the new code of Section V-B will have ML decoding complexity less than or equal to that of the code generated by using the $N$ antenna design from $\mathcal{F}_{2,AG}$ as the base. It is straightforward to show that for $N > 5$, the number of real symbols in the design from $\mathcal{F}_{2,AG}$, say $K'_b$, satisfies $K'_b > 2N$. From (11), the new code gives lower decoding complexity if

$$M^{\frac{K-K_b}{2}} M^{\frac{K_b}{4}-\frac{1}{2}} > M^{\frac{K-K'_b}{2}} M^{\frac{K'_b}{4}-\frac{1}{2}}.$$

This condition is equivalent to $K_b < K'_b$. Since this is indeed the case, we have proved the required result. ∎

*3) Comparison with other codes:* In [27], FD codes, called Embedded Alamouti Space-Time (EAST) codes, were constructed for rates $R = 1, 2, \ldots, N/2$ cspcu for $N = 2, 4, 6$ and 8 antennas. Table III shows that the new codes of Section V-B have lower ML decoding complexity in each case. The EAST codes, however, have non-vanishing determinant property. In [5], an FGD code for 4 antennas and rate $17/8$ was constructed and this code was punctured to obtain a rate 2 code for 4 antennas. From Table III, we see that the new codes of Section V-B have lower ML decoding complexity than the codes in [5]. In [28], a rate $3/2$ cspcu, 4 antenna code was constructed with complexity $M^3$ (see Examples 5 and 14). The new code of section V-B has complexity only $M^{2.5}$. The rate 2 code for 4 antennas constructed in [29] has an ML decoding complexity of order $M^6$ and non-vanishing determinant property. The new code that we report for $N = 4$ and $R = 2$ has a considerably lower ML decoding complexity of the order of $M^{4.5}$. The new codes match the code from [22] in decoding complexity for $N = 4$, $R = 2$ and have lower decoding complexity for $N = 2$, $R = 2$. The new $N = 2$, $R = 2$ code has same decoding complexity as the Silver code [25], [26] and the new code has lower ML decoding complexity than the Golden Code [24].

From the above discussion it is clear that for $N \leq 5$ and $R > 1$ the complexity of ML decoding the new codes of Section V-B match the least known ML decoding complexity in the literature. When $N \geq 6$, $N$ not a power of 2, and $R > 1$, the new codes have lower ML decoding complexity than the codes available in the literature. When $N \geq 8$, $N$ a power of 2, and $R > 3/2$, again the new codes have lower ML decoding complexity than already known codes. When $N \geq 8$, a power of 2 and $1 < R \leq 3/2$, the new codes either have lower ML decoding complexity or match the least known ML decoding complexity in the literature. Thus, the subset of the new codes from Section V-B corresponding to $R = N$ are asymptotically-optimal, fast-decodable and have least ML decoding complexity among all known full-rate, full-diversity codes in the literature.

## VI. Discussion

In this paper, we have constructed new classes of asymptotically-good, full-diversity $g$-group ML decodable codes, $g > 1$ and new FGD codes. Using these codes we have constructed STBCs of least known ML decoding complexity for $N > 1$ and $R > 1$ which include a class of asymptotically-optimal full-diversity, fast-decodable codes. The results of this paper lead us to the following questions.

- Determining the highest possible rate of full-diversity, multigroup ML-decodable codes.
- Is there an algebraic theory behind the constructions of multigroup ML decodable codes given in this paper? Are there other high-rate multigroup ML-decodable codes? Can improvements be made on rates reported in this paper?
- Is it possible to derive a lower bound on the ML decoding complexity of the class of full-diversity STBCs of a given size $N$ and rate $R$?
- How can one explicitly construct signal sets, that provably lead to full-diversity and large coding gain, given a design with full-rank weight matrices?
- Proving the non-vanishing determinant property, either in the affirmative or otherwise, of the new class of asymptoically-optimal, fast-decodable codes proposed in this paper remains an interesting direction to pursue.